\documentclass[11pt]{article}
\usepackage{graphicx,color}
\usepackage{appendix}
\usepackage{latexsym,amsmath,amssymb,graphicx,booktabs}
\usepackage{graphicx,amsmath,amsfonts,amssymb,slashed,upgreek,caption,amscd,cancel, tensor,color}
\usepackage{adjustbox}
\usepackage{epsfig,latexsym,cite}
\usepackage{hyperref}
\numberwithin{equation}{section}

\definecolor{MyBlue}{rgb}{0.15,0.15,0.70}

\hypersetup{
colorlinks=true,
citecolor=MyBlue,
linkcolor=MyBlue,
urlcolor=MyBlue
}

\input epsf
\usepackage{amssymb}
\usepackage{amsmath}
\usepackage{amsfonts}
\usepackage{upgreek}
\usepackage{latexsym}

\newcommand{\be}{\begin{equation}}
\newcommand{\ee}{\end{equation}}
\newcommand{\beq}{\begin{equation}}
\newcommand{\eeq}{\end{equation}}
\newcommand{\bea}{\begin{eqnarray}}
\newcommand{\eea}{\end{eqnarray}}

\newcommand{\DD}{\hat{\alpha}}
\newcommand{\DDt}{\alpha}

\newcommand{\sR}{R}
\newcommand{\B}{{\cal B}}

\newcommand{\F}{{\cal F}}
\newcommand{\Q}{{\cal Q}}

\usepackage[stable]{footmisc}
\def\d{\delta}

\def\dkmu2{\delta K_{\mu \nu}\delta K^{\mu \nu}}
\def\pmu2{  \phi_{\mu \nu}\phi^{\mu \nu}}

\newcommand{\nbeta}{\beta}
\newcommand{\mbeta}{\beta}
\newcommand{\bs}{{\bar \sigma}}
\newcommand{\bQ}{{\bar Q}}

\newcommand{\LS}{{{\cal A}}_K}
\newcommand{\A}{\hat{\cal A}_K}
\newcommand{\Ch}{{{\cal C}}}
\newcommand{\Co}{\hat{\cal C}}
\newcommand{\LRR}{\hat{\cal A}_R}
\newcommand{\LZ}{{{\cal A}}_R}

\newcommand{\LRo}{{\cal G}}
\newcommand{\LNRo}{{\cal B}_R}

\newcommand{\bN}{\bar N}

\newcommand{\tg}{\check{g}}
\newcommand{\NS}{{N_S}}

\renewcommand\[{\left[}

\newcommand\ees{\end{eqnarray}}
\newcommand\bees{\begin{eqnarray}}

\newcommand\alphaDI{\alpha_{\text{D},I}}
\newcommand\alphaCI{\alpha_{\text{C},I}}
\newcommand\alphaDb{\alpha_{\text{D},b}}
\newcommand\alphaCb{\alpha_{\text{C},b}}
\newcommand\alphaDc{\alpha_{\text{D},c}}
\newcommand\alphaCc{\alpha_{\text{C},c}}
\newcommand\alphaD{\alpha_{\text{D}}}
\newcommand\alphaC{\alpha_{\text{C}}}
\newcommand\alphaB{\alpha_{\text{B}}}
\newcommand\alphaM{\alpha_{\text{M}}}
\newcommand\alphaK{\alpha_{\text{K}}}
\newcommand\alphaT{\alpha_{\text{T}}}
\newcommand\cI{c_{s,I}}
\newcommand\tcI{\tilde c_{s,I}}
\newcommand\tbeta{\beta}
\newcommand\gammac{\gamma_c}
\newcommand\rhoD{\rho_{\rm DE}}
\newcommand\pD{p_{\rm DE}}
\newcommand\wD{w_{\rm DE}}
\newcommand\kH{k_H}

\begin{document}

\begin{center}
\Large{\textbf{Effective  Theory of Interacting Dark Energy  
}} \\[1cm] 
 
\large{J\'er\^ome Gleyzes$^{\rm a,b}$,  David Langlois$^{\rm c}$, \\[0.1cm]
Michele Mancarella$^{\rm a,b}$ and Filippo Vernizzi$^{\rm a}$}
\\[0.5cm]

\small{
\textit{$^{\rm a}$ CEA, IPhT, 91191 Gif-sur-Yvette c\'edex, France \\ [0.05cm]
CNRS,  URA-2306, 91191 Gif-sur-Yvette c\'edex, France}}

\vspace{.2cm}

\small{
\textit{$^{\rm b}$ Universit\'e Paris Sud, 15 rue George Cl\'emenceau, 91405,  Orsay, France}}

\vspace{.2cm}

\small{
\textit{$^{\rm c}$  APC, (CNRS-Universit\'e Paris 7), 10 rue Alice Domon et L\'eonie Duquet, 75205 Paris, France \\ 
}}
\vspace{.2cm}

\vspace{0.5cm}
\today

\end{center}

\vspace{2cm}

\begin{abstract}
We present a unifying treatment of dark energy and modified gravity that allows distinct conformal-disformal couplings of matter species to the gravitational sector. In this very general approach, we derive the conditions to avoid ghost and gradient instabilities.  We compute the equations of motion for background quantities and linear perturbations. We illustrate our formalism with two simple scenarios,  where either cold dark matter or  a relativistic fluid is nonminimally coupled.  This extends previous studies of coupled dark energy to  a much broader spectrum of  gravitational theories.
\end{abstract}

\newpage 
\tableofcontents

\vspace{.5cm}
\newpage
\section{Introduction}

The nature of dark energy, responsible for the present cosmological acceleration, is a central topic in theoretical and observational cosmology. 
One of the main goals of current and future cosmic surveys 
is to constrain or possibly detect deviations from the standard $\Lambda$CDM scenario, induced by the presence of dark energy or modifications of General Relativity (GR) (see e.g.~\cite{Amendola:2012ys}).  This is particularly relevant on scales above $\sim 10$ Mpc, where deviations from GR are not yet well tested. Fortunately, on these scales cosmological perturbations are still in the linear regime today and linear perturbation theory around a FLRW background is thus a valid description. 

Given the plethora of existing dark energy and modified gravity models (see for instance \cite{Clifton:2011jh,Joyce:2014kja}), it is worth resorting to an effective approach that tries to describe all possible deviations from $\Lambda$CDM  in a simple and systematic way, relying on a minimal number of parameters. In the linear regime for perturbations, this task has been sucessfully undertaken  for single scalar field models  in \cite{Gubitosi:2012hu,Gleyzes:2013ooa}. Initially inspired by the so-called Effective Field Theory of inflation \cite{Creminelli:2006xe,Cheung:2007st} and minimally coupled dark energy \cite{Creminelli:2008wc}, this approach relies on the construction of an effective action for linear perturbations. In order to do so, we start from a generic Lagrangian written in terms of Arnowitt-Deser-Misner (ADM) \cite{Arnowitt:1962hi} quantities defined with respect to the uniform scalar field hypersurfaces (see also \cite{Bloomfield:2012ff,Bloomfield:2013efa} for an analogous approach, \cite{Piazza:2013coa,Tsujikawa:2014mba,Gleyzes:2014rba} for recent reviews and e.g.~\cite{Bloomfield:2013cyf,Piazza:2013pua,Bellini:2014fua} for applications). After having been  implemented in a public numerical code named EFTCAMB \cite{Hu:2013twa},
most recently, it has been  applied to constrain deviations from the standard cosmological constant scenario by the Planck collaboration \cite{Ade:2015rim}.

The action developed in \cite{Gleyzes:2013ooa} contains five free functions of time that parametrize any deviation from $\Lambda$CDM. Four of these functions describe cosmological perturbations in Horndeski theories \cite{Horndeski:1974wa,Deffayet:2011gz,Kobayashi:2011nu}. The fifth parameter describes deviations from GR encompassing Horndeski theories. Indeed, the same formalism was also instrumental to uncover the theories beyond Horndeski of~\cite{Gleyzes:2014dya,Gleyzes:2014qga}, which lead to equations of motion higher than second order but are free from Ostrogradski instabilities (see e.g. \cite{Zumalacarregui:2013pma} for an earlier example of theories beyond Horndeski).

The developments described above assume that  matter is  minimally coupled to a unique metric, which will be called Jordan frame metric for convenience. 
However, although the universality of couplings is very well tested  on Solar System scales \cite{Will:2014xja}, on cosmological scales constraints are much weaker. In particular, the scalar field  responsible for the current accelerated expansion is known to mediate a fifth force \cite{Brans:1961sx}, which may lead to violations of the equivalence principle (EP) on large scales \cite{Hui:2009kc}  (see also \cite{Creminelli:2013nua} for a test of the EP on large scales). Moreover, while fifth force effects on standard matter such as baryons and photons are severely restricted, those on cold dark matter (CDM) or neutrinos could be much larger. This leaves the freedom to consider the case where different matter species\footnote{By matter species we intend the different components in the Universe (baryons, photons, CDM and neutrinos) but the results derived here could be straightforwardly  extended to  different types of objects, such as e.g.~galaxies of different sizes, behaving differently under the effect of the fifth force.} couple differently to the scalar field \cite{Damour:1990tw,Amendola:1999er}.

The goal of the present work is to extend the approach developed in Refs.~\cite{Gubitosi:2012hu,Gleyzes:2013ooa} by relaxing the assumption that all matter species are minimally coupled to the same metric.\footnote{Another general parametrisation of theories of single-field dark energy that is explicitly coupled to dark matter has been given in \cite{Skordis:2015yra}, in the framework of the Parameterized Post-Friedmann approach \cite{Baker:2012zs,Ferreira:2014mja}.}  
For simplicity, in the following we restrict our study to effective theories  of dark energy or modified gravity that remain within the  Horndeski class. This means that  we assume that the function $\alpha_{\rm H}$ introduced in \cite{Gleyzes:2014qga} to describe theories beyond Horndeski at the level of linear perturbations vanishes here, leaving only four out of the five free independent functions of \cite{Gleyzes:2013ooa}. We reserve  a treatment of theories beyond Horndeski for future work.
As shown in \cite{Bettoni:2013diz}, the structure of  the Horndeski Lagrangians is preserved under a disformal transformation \cite{Bekenstein:1992pj}  of the metric with coefficients that depend only on the scalar field (not on its gradient), i.e.~of the form 
\be
\label{disf_unit}
\tilde{g}_{\mu \nu} = C(\phi) g_{\mu \nu}  + D(\phi) \partial_\mu \phi \, \partial_\nu \phi \;.
\ee
Thus, in the following we  assume that each matter species is minimally coupled to a distinct Jordan metric of this form.\footnote{Other types of couplings can be found in the literature. For instance, Ref.~\cite{Blas:2012vn}  considers a CDM action that  depends on the contraction of the CDM 4-velocity with the normalized space-time gradient of the scalar field, in the context of Lorentz-violating theories. Ref.~\cite{Bettoni:2015wla} directly modifies the action for a general perfect fluid.}
While conformal couplings (i.e.~with $D=0$) have been extensively studied in the literature (see e.g.~\cite{Amendola:2012ys} and references therein), disformal couplings have been investigated only recently (see also e.g.~\cite{Koivisto:2008ak,Zumalacarregui:2010wj,Koivisto:2012za,vandeBruck:2012vq,Zumalacarregui:2012us,Brax:2013nsa,Brax:2014vva,Sakstein:2014isa,vandeBruck:2015ida,Koivisto:2015mwa}). Moreover, the dynamics of the gravitational metric $g_{\mu\nu}$ is usually assumed to be governed by the standard Einstein-Hilbert action. Here, we allow a much more general gravitational sector, based on the effective description given in~\cite{Gleyzes:2013ooa}.  
In Sec.~\ref{section2} we review our formalism within the ADM effective approach and  the gravitational action in the uniform scalar-field gauge. Apart from the four time-dependent parameters mentioned above, we introduce  
two extra functions of time for each species, describing the  nonminimal coupling to dark energy via an effective metric of the form \eqref{disf_unit}. 
The structure of this action  is preserved under  transformations of the reference metric of the form \eqref{disf_unit} and the stability conditions for the matter {\em and} the gravitational sector are shown to be invariant under these transformations. More details on the frame dependence and on the derivation of the stability conditions of gravitational and matter quantities are respectively given in Appendix \ref{app2} and Appendix \ref{Lagmatter}. 

In Sec.~\ref{sec3} we derive the evolution equations describing the matter sector, which now include the effect of the  nonminimal couplings, and in Appendix \ref{parameters} we provide the definitions of several parameters introduced in this section. These equations must be supplemented with the Einstein equations describing the gravitational sector, reported in Appendix~\ref{app1}.  We provide and discuss the perturbation equations  using Newtonian gauge but these are also given in synchronous gauge in Appendix \ref{app_sync}.

The parameters of our effective description can be constrained by observations. As a direct application of our approach, in Sec.~\ref{sec4} we consider the cosmological consequences, for the background evolution and for linear perturbations, of a Universe where the coupling of CDM differs from that of the other species (see e.g.~\cite{Billyard:2000bh,Farrar:2003uw,Amendola:2003wa,Bertolami:2007zm,Baldi:2008ay,Koyama:2009gd,Valiviita:2009nu,Baldi:2010vv,Pettorino:2012ts,yang:2014vza,Salvatelli:2014zta}). Our analysis extends previous results as we allow gravity itself to be modified, not only the couplings to matter. In Sec.~\ref{sec5} we consider the case where the coupled species is a relativistic fluid. This will allow us to highlight the dependence of conformal and disformal couplings on the equation of state. Finally, we conclude in Sec.~\ref{sec_last}.


\section{Unifying description of dark energy with non universal couplings}
\label{section2}

In this section we introduce the gravitational and matter actions within the ADM framework introduced in 
\cite{Gleyzes:2013ooa} and recently summarized in \cite{Gleyzes:2014rba}. After giving the background equations of motion, we study linear fluctuations and derive the conditions for the linear theory to be healthy, i.e.~ghost-free and without gradient instabilities.

\subsection{Gravitational and matter actions}
\label{sec2}

In the present work, we assume that the gravitational sector is described by a four-dimensional metric $g_{\mu \nu}$ and a scalar field $\phi$. 
Let us start by choosing a coordinate system such that the constant time hypersurfaces coincide with the  uniform scalar field hypersurfaces. 
In this gauge, referred to  as unitary gauge, the metric  can be written in the ADM form, which   reads
  \be
\label{ADM}
ds^2=-N^2 dt^2 +{h}_{ij}\left(dx^i + N^i dt\right)\left(dx^j + N^j dt\right) \, ,
\ee
where $N$ is the lapse and $N^i$ the shift.  In the following,  a dot will stand for a time derivative with respect to $t$, and $D_i$ will denote the covariant derivative associated with the three-dimensional spatial metric $h_{ij}$. Spatial indices will be lowered and raised with the spatial metric $h_{ij}$ or its inverse $h^{ij}$, respectively.

In the unitary gauge, a generic gravitational action  can be written in terms of  geometric quantities that are invariant under spatial diffeomorphisms \cite{Creminelli:2006xe,Cheung:2007st}. Expressed in ADM coordinates introduced above, these geometric quantities are the lapse $N$, the extrinsic curvature $K_{ij}$ of the constant time hypersurfaces, whose components are given by
 \be
\label{extrinsic_ADM}
K_{ij} = \frac{1}{2N} \big(\dot h_{ij} - D_i N_j - D_j N_i \big) \;,
\ee
as well as  the 3d Ricci tensor $R_{ij}$ of the constant time hypersurfaces  and, possibly,  spatial derivatives of all these quantities. Thus, the action is generically of the form
 \beq
\label{g_action}
S_{\rm g}=\int d^4 x \sqrt{-g}\,  L(N,  K_{ij}, R_{ij}, h_{ij},D_i ;t) \;.
\eeq

The gravitational action must be supplemented by a matter action $S_{\rm m}$.
In order to describe dark energy and modified gravity scenarios with EP violations, we  assume that beside the gravitational sector introduced above, the Universe is filled by $\NS$ matter species labelled by an index $I$, with $I=1, \ldots, \NS$, each minimally coupled to a different metric. For each species $I$, we denote the corresponding metric by  $\tg^{(I)}_{\mu \nu}$ and we call this the Jordan frame metric associated with this species.
The total matter action is thus given by 
\be
S_{\rm m} = \sum_I^\NS S_I \;, \qquad S_I =  \int d^4 x \sqrt{- \tg^{(I)} }\,  L_I \Big(   \tg^{(I)}_{\mu \nu}, \psi_I \Big) \; ,
\ee
with
\be
\label{disf_unit_I2}
\tg^{(I)}_{\mu \nu} = C^{(\phi)}_I(\phi) g_{\mu \nu}  + D^{(\phi)}_I(\phi) \partial_\mu \phi \, \partial_\nu \phi \;.
\ee
In order to preserve the Lorentzian signature of the Jordan-frame metric of the species $I$, it is necessary to have  $C_I^{(\phi)}>0$.

There is some arbitrariness in the choice of the gravitational metric $g_{\mu\nu}$ since we work in the context of modified gravity, where the gravitational dynamics cannot be expressed in terms of a standard Einstein-Hilbert term, in general. It is  often convenient to choose one particular matter species, say $I_*$, and define its Jordan metric as the gravitational metric, in which case we have $C^{(\phi)}_{I_*}=1$ and $D^{(\phi)}_{I_*}=0$. 

\subsection{Homogeneous equations}
\label{sec:hom_1}

Let us discuss briefly the  evolution of the background metric described by a FLRW metric assumed to be spatially flat. In this case the lapse is a  function of time only, which we denote $\bar N(t)$, the shift vanishes, $N^i=0$,  and  the spatial metric reads $g_{ij} =h_{ij}= a^2(t) \delta_{ij}$ where $a$ represents  the scale factor. Thus, the metric reads
\be
\label{FLRW}
ds^2 = - \bar N^2(t) dt^2 + a^2(t) d \mathbf{x}^2 \;.
\ee

The homogeneous dynamics depends on the gravitational Lagrangian $L$ in eq. (\ref{g_action}), which can be seen as a function $\bar{L}(N, a, \dot a)$ when the arguments are restricted to their 
 background values, i.e. $N=\bar{N}$, $h_{ij}= a^2(t) \delta_{ij}$, $R_{ij}=0$, and 
 \be
 K_{ij}=\bar{K}_{ij}\equiv \frac{a \dot a}{\bar N} \delta_{ij}=H h_{ij}\,,
 \ee
 where $H\equiv \dot a /(a \bN)$ denotes the Hubble rate. Here and in the following, barred quantities are evaluated on the background.
 
The  variation of the matter action $S_{\rm m}$  with respect to the metric $g_{\mu\nu}$ defines the energy-momentum tensor, according to  the standard expression
 \be
T^{\mu\nu} \equiv  \frac{2}{\sqrt{-g}} \frac{\delta S_{\rm m}}{\delta g_{\mu\nu}} \;. 
\ee
This definition applies even if the matter is minimally coupled with respect to a metric $\tg_{\mu\nu}$ that differs from $g_{\mu\nu}$, as discussed in Appendix \ref{app2}. In the homogenous case, the energy-momentum tensor depends only on the energy density $\rho_{\rm m} \equiv -\bar T^{0}_{\ 0}$ and  the pressure $p_{\rm m} \equiv \bar T^i_{\ i}/3$. If there are several matter components, the previous quantities  simply correspond, respectively, to the sums of the energy densities and pressures associated to each individual species, i.e.~$\rho_{\rm m} = \sum_I \rho_I $ and $p_{\rm m} = \sum_I p_I $.

The background evolution equations are then obtained by taking the variation of the total homogeneous action $S_{\rm g} + S_{\rm m}$ with the respect to $\bar N$ and $a$. As shown in \cite{Gleyzes:2013ooa}, this leads to the equations 
\be
 \label{Fried1}
 \bar L+\bN L_N-3H \F=\rho_{\rm m}
 \ee
 and 
 \be
 \label{Fried2}
 \bar L-3H\F-\frac{\dot \F}{\bN}=- p_{\rm m}\,,
 \ee
  where  the coefficient $\F$ is defined by
 \be
 \left(\frac{\partial L}{\partial K_{ij}}\right)_{\rm bgd}\equiv \F\,  h^{ij}\,.
 \ee
Equations~\eqref{Fried1} and~\eqref{Fried2} generalize the usual Friedmann equations. 
For GR, where  the Lagrangian is given by $L=M_P^2(K_{ij}K^{ij}-K^2+R)/2$, one can check that the standard equations are recovered, since $\bar L=-3M_P^2 H^2$, $L_N=0$ and $\F=-2M_P^2 H$. 

 The Friedmann equations eqs.~\eqref{Fried1}--\eqref{Fried2} can  always be written as 
\begin{align}
H^2 &= \frac{1}{3 M^2} \left(  \rho_{\rm m}  + \rhoD \right)  \;, \label{Fr1} \\
 \dot H + \frac32 H^2 &= - \frac{1}{2M^2 }\left(  p_{\rm m} +  \pD \right) \label{Fr2} \;,
\end{align} 
where $M$ denotes the effective Planck mass, which can be  in general time-dependent (it will be defined below from the second derivative of $L$ with respect to the intrinsic curvature). The above equations can be interpreted as  definitions of the homogeneous energy density and pressure of dark energy, respectively given by 
\be
 \rhoD \equiv  3 M^2 H^2  - \rho_{\rm m} \;, \qquad \pD \equiv - M^2 (2\dot H + 3 H^2) - p_{\rm m} \;.
 \ee 
These  equations can also be shown to be equivalent to the Friedmann equations derived from 
the Lagrangian \cite{Gubitosi:2012hu,Gleyzes:2013ooa}
\be 
L = \frac{M^2}{2} {}^{(4)}\!R + \frac{c}{N^2} - \Lambda  \;, \label{action_back}
\ee
where ${}^{(4)}\!R$ is the 4d Ricci scalar and $c=c(t)$ and $\Lambda=\Lambda(t)$ are time-dependent functions, respectively given by
\begin{align}
2 c  & = \rhoD + \pD + H (M^2)^{\hbox{$\cdot$}} -(M^2)^{\hbox{$\cdot \cdot$}} \;, \\ 
2 \Lambda & = \rhoD -\pD + 5 H (M^2)^{\hbox{$\cdot$}} + (M^2)^{\hbox{$\cdot \cdot$}}  \;.
\end{align}

 \subsection{Linear perturbations} 
We  now expand the gravitational action up to second order in perturbations, in terms of the perturbative quantities 
\be
\delta N = N -\bN(t)\,, \qquad \delta K_{ij} = K_{ij} - H h_{ij}\, ,
\ee
as well as $R_{ij}$, which is already a perturbation since its background value vanishes.

The second-order expansion of the gravitational Lagrangian involves first and second derivatives of $L$ with respect to its arguments $K_{ij}$, $R_{ij}$ and $N$. It is convenient to introduce
 the time-dependent coefficients $\LRo$, $\LNRo$, $\B$, $\A$, $\LS$, $\hat {\cal C}$, ${\cal C}$, $\LRR$ and $\LZ$ respectively as
 \be
 \frac{\partial L}{\partial R^i_j}=\LRo \, \d_i^j \;, \qquad \frac{\partial^2 L}{\partial N\partial R^i_j}={\LNRo}\,\d_i^j \;, \qquad \frac{\partial^2 L}{\partial N\partial K^i_j} =\B \, \d_i^j\,,
\ee
\begin{align}
\label{AAA}
\frac{\partial^2 L}{\partial K_i^j\,\partial K_k^l} & =\A \,  \d^i_j \, \d^k_l+ \LS\left(\d^i_l\, \d^k_j+\d^{ik} \d_{jl}\right)\,, \\
\frac{\partial^2 L}{\partial R_i^j\, \partial R_k^l} & =\LRR\,  \d^i_j \, \d^k_l+ \LZ \left(\d^i_l\, \d^k_j+\d^{ik} \d_{jl}\right)\,, \\
\frac{\partial^2 L}{\partial K_i^j\, \partial R_k^l} &= \Co\,  \d^i_j \, \d^k_l+ \Ch \left(\d^i_l\, \d^k_j+\d^{ik} \d_{jl}\right)\,,
\end{align}
where all partial derivatives on the left hand sides are evaluated on the background. The form of the right hand side of these expressions is merely determined by the FLRW symmetries. The first and second derivatives of $L$ with respect to the scalar $N$ are simply denoted as $L_N$ and $L_{NN}$, respectively.

In the following, for simplicity, we restrict our considerations to Lagrangians that lead to dynamical equations with at most  two space derivatives. This is automatically ensured if we impose the conditions \cite{Gleyzes:2013ooa,Gleyzes:2014rba}\footnote{Here we have corrected a typo in Ref.~\cite{Gleyzes:2014rba}. The coefficient in front of $\delta K \delta R$ inside the bracket in eq.~(55) (see v2 of the arXiv version) should be $\hat{\cal C}/2$, so that the condition in the second line of eq.~(60)  should read $\hat{\cal C}^*=\hat{\cal C}+\cal C$. With this correction,  eq.~(76) of Ref.~\cite{Gleyzes:2014rba} is equivalent to eq.~\eqref{conds} in this article.}
\beq
\label{conds}
\A+2\LS =0\;, \qquad  \hat {\cal C} +  {\cal C} =0 \; ,\qquad 4\LRR+3\LZ=0\; .
\eeq
We also  impose  the further condition 
\be
\LNRo = \frac{1}{\bN} (\LS -  \LRo - H {\cal C}) \; ,
\ee
which is equivalent to restricting the range of application of the expanded action to Horndeski theories \cite{Gleyzes:2013ooa}.\footnote{To parametrize deviations from Horndeski theories at the linear level, Ref.~\cite{Gleyzes:2014qga} introduced the parameter $\alpha_{\rm H} \equiv ({\cal G} + H {\cal C}+ \bar N {\cal B}_R )/{\cal A}_K - 1$. Here we will assume $\alpha_{\rm H} =0$.}

The second-order gravitational action can then be explicitly written in terms of all the coefficients introduced above. In fact, the quadratic action involves only a few combinations of these coefficients, which are represented by the following dimensionless parameters 
 \cite{Bellini:2014fua,Gleyzes:2014rba}
\be
 \alphaK \equiv \frac{2 \bN L_N + \bN^2 L_{NN}}{2 H^2 \LS} \;, \qquad \alphaB \equiv \frac{\B \bN}{4 H \LS} \;, \qquad \alphaT \equiv \frac{{\LRo}+{\dot\Ch}/{(2\bar{N})} + H{\cal C}}{\LS} -1 \;.
\ee
The  effective Planck mass squared  is defined by $M^2 \equiv 2 \LS$. With this definition, $M$ coincides with the time-dependent Planck mass introduced in eqs.~\eqref{Fried1} and  \eqref{Fried2} and in the action \eqref{action_back}.
Its possible time variation is characterized by
\beq
\label{alphaM}
\alphaM\equiv \frac{1}{\bN H}\frac{d \ln M^2}{ dt}\,.
\eeq
In terms of these parameters, one finds that 
the second-order gravitational action is given by \footnote{To write this action, we have not assumed $\bN =1$ as done in previous references \cite{Gubitosi:2012hu,Gleyzes:2013ooa,Gleyzes:2014rba}. In such a way the action remains explicitly invariant under a time reparameterization $t \to \tilde t(t)$, which is convenient when changing frame.}
\be
\begin{split} 
\label{S2}
 S_{\rm g}^{(2)}=  \int d^3x dt \,a^3  \bN \frac{M^2}{2}   \bigg[ & \delta K^i_j \delta K^j_i-\delta K^2 + R \frac{\delta N}{\bN}+(1+\alphaT) \, \delta_2 \Big(   {\sqrt{h}}R/{a^3 }\Big)    \\
 &   +  \alphaK H^2 \left(\frac{\delta N}{\bN}\right)^2 + 4 \alphaB H \delta K \frac{\delta N}{\bN}  \bigg]   \, ,
 \end{split}
 \ee
where $\delta_2$ denotes taking the expansion at second order in the perturbations. Moreover, we have omitted irrelevant terms that vanish when adding the matter action and imposing the background equations of motion.

To verify that $M$ plays the role of the Planck mass which canonically normalizes the graviton, let us write this action  in terms of the tensor fluctuations, defined as the traceless and divergence-free fluctuations of the spatial metric, i.e. 
\be
h_{ij} = a^2(t) \left(\delta_{ij} + \gamma_{ij}\right) \;, \qquad \gamma_{ii}=0 = \partial_i \gamma_{ij} \;.
\ee
The above action then yields
\be
S_{\gamma}^{(2)} =\int dx^3 dt \, a^3  \frac{M^2}{8 \bar N} \left[   \dot{\gamma}_{ij}^2 -  c_T^2 \frac{\bar N^2}{a^2}(\partial_k \gamma_{ij})^2 \right] \;,
\ee
where the tensor sound speed squared is given by $c_T^2 \equiv   1 + \alphaT$.
Absence of ghosts and gradient instabilities respectively require that the  kinetic and  spatial gradient terms  are positive, i.e.~that
\be
M^2 \geq 0 \;, \qquad \alphaT \ge -1 \;,
\ee
which will be assumed in the following.

\subsection{Matter couplings and stability conditions}
\label{sec:ma}

To discuss the stability and determine the propagation speed of dark energy perturbations,  one must also include quadratic terms  that come from the matter action, because the latter depends on the gravitational degrees of freedom. In order to do so, we need to to take into account the fact that  each matter species $I$ is minimally coupled to a metric $\tg_{\mu \nu}^{(I)}$ defined in eq.~\eqref{disf_unit_I2}. For later convenience, we define, for each matter species, the time-dependent quantity
\be
\label{alphaCphi}
\alphaCI \equiv \frac{\dot{\phi}}{2 H N} \frac{d \ln C_I^{(\phi)} }{d\phi}\;,  
\ee
which parameterizes how the conformal coupling affects physical observables;
the impact of the disformal coupling is parameterized by the quantity\footnote{This parameter coincides with $1/\gamma^2$, where $\gamma$ is the so-called disformal scalar in the notation of Ref.~\cite{vandeBruck:2015ida}.}
\be
\label{alphaDphi}
 \alphaDI \equiv \frac{ (\dot{ \phi}/ N)^2 D_I^{(\phi)}}{ C_I^{(\phi)} -(\dot{ \phi}/N)^2  D_I^{(\phi)} } \;,
\ee
and the right-hand side of these equations  are to be evaluated on the background. Requiring that the Jordan frame metric is Lorentzian implies $\alphaDI > -1$  \cite{Bettoni:2013diz}.

In unitary gauge, eq.~\eqref{disf_unit_I2}  reads
\be
\label{disf_unit_I}
\tg^{(I)}_{\mu \nu} = C_I(t) g_{\mu \nu}  + D_I(t) \delta_\mu^0 \delta_\nu^0 \;,
\ee
with
\be
C_I(t) =  C_I^{(\phi)} \big( \phi( t) \big)\, , \qquad D_I(t) =   \dot{ \phi}^2 (t) D_I^{(\phi)} \big( \phi( t) \big)\,.
\ee
Then the parameters $\alphaCI$ and $\alphaDI$ introduced above take the form
\be
\label{alphaCD}
\alphaCI = \frac{1}{2 H \bN} \frac{d \ln C_I}{dt}\, , \qquad  \alphaDI = \frac{D_I}{ \bN^2 C_I-D_I} \;. 
\ee

Combining the quadratic action for matter with eq.~\eqref{S2}, one can extract a quadratic action that governs the dynamics of the gravitational scalar degree of freedom and the matter ones. The explicit calculation in the case of perfect fluids is presented in Appendix~\ref{Lagmatter}. The absence of ghosts is guaranteed by the positivity of the matrix in front of the kinetic terms. For the gravitational scalar degree of freedom, this condition is given by

\be
\DDt \equiv \alphaK + 6 \alphaB^2 +3 \sum_I \alphaDI \, \Omega_I \geq0\;, \label{alpha_def}
\ee
 where we have introduced the  (time-dependent) dimensionless density  parameter
\be
\Omega_I\equiv \frac{\rho_I}{3M^2 H^2}\,,
\ee
where we recall that $M^2$ is in general time dependent. As pointed out already in\cite{Bruneton:2007si,Zumalacarregui:2012us}, the presence of a disformal coupling affects the ghost-free condition. 

For the matter sector,  the analogous condition usually corresponds to the Null Energy Condition \cite{Hawking:1973uf}.  
In the Jordan frame of each species $I$, this can be expressed in terms of the energy density and pressure by $\check{\rho}_I+\check p_I \geq 0 $ (we use the  symbol $\check{}$ to denote Jordan-frame quantities). In the frame of $g_{\mu \nu}$, this inequality becomes
\be
\label{NECdisf}
\rho_I+(1+\alphaDI)p_I \geq 0 \, ,
\ee
where we have used that $\check{w}_I = (1+\alphaDI)w_I$ (see Appendix~\ref{app2} for various  relations between quantities defined in distinct frames).

The speed of sound for  scalar perturbations can be read off from the quadratic action derived in Appendix~\ref{Lagmatter}. One finds
\be
\label{ss}
c_s^2 = -\frac{2}{\DDt}\bigg\{ (1+\alphaB) \bigg[ \frac{\dot H}{H^2} -\alphaM + \alphaT  +  \alphaB (1+\alphaT)\bigg] + \frac{\dot \alpha_{\rm B}}{H} +\frac32 \sum_I  \Big[ 1+ (1+\alphaDI) w_I \Big] \Omega_I\bigg\} \;,
\ee
where matter appears in the last term in the bracket, proportional to $ \sum_I  (\check{\rho}_I+\check p_I) $. Absence of gradient instabilities is guaranteed provided that
\be
c_s^2\geq 0\,.
\ee
We also require that the propagation speed for each matter species, in its  Jordan frame, is positive, $ \check c_{s,I}^2\geq0 $.

\subsection{Disformal transformations}
\label{sec:disf}
As mentioned earlier, there is  some arbitrariness in the choice of the metric $g_{\mu\nu}$ that describes the gravitational sector. Let us thus see how the description is modified when the reference metric undergoes a disformal transformation, of the form
\be
g_{\mu \nu} \to \tilde g_{\mu \nu} = C^{(\phi)}( \phi ) g_{\mu \nu} +  D^{(\phi)} ( \phi ) \partial_\mu \phi \partial_\nu \phi \;. \label{disf_trans}
\ee
In unitary gauge,  this corresponds to the transformation
\be
\label{disf_uni}
g_{\mu \nu} \to \tilde g_{\mu \nu} = C(t) g_{\mu \nu}  + D(t) \delta_\mu^0 \delta_\nu^0  \;,
\ee
with $C(t) =  C^{(\phi)} \big( \phi( t) \big)$ and $D(t) =  D^{(\phi)} \big( \phi( t) \big) \dot \phi^2 (t)$.
The effect of this transformation on the ADM quantities, on the background quantities and on the linear perturbations is described in detail in Appendix \ref{app2}. Here, we just present the main consequences on the parametrization of the couplings and of the linear perturbations. 

In analogy with \eqref{alphaCD}, it is convenient to introduce the dimensionless time-dependent parameters 
 \be
\label{defalphas}
 \alphaC \equiv \frac{\dot C}{2 H \bN C} \, , \qquad \alphaD \equiv \frac{D}{ \bN^2 C-D}\, ,
 \ee
 which characterize, respectively, the conformal and disformal parts of the above metric transformation.\footnote{We require $C>0$ and $\alphaD > -1$, see discussions respectively in Secs.~\ref{sec2} and \ref{sec:ma}.}
 
Let us first see how  the gravitational  action \eqref{S2} changes under the transformation (\ref{disf_uni}). 
As shown in Ref.~\cite{Bettoni:2013diz}, the structure of Horndeski Lagrangians is preserved under a disformal transformation. Indeed, using eqs.~\eqref{disformalADM} and \eqref{disformalRK}, one can check that  \eqref{S2} maintains the same structure with 
the time-dependent coefficients in the action transforming as
\be
 \tilde{M}^2 =\frac{M^2 }{ C\sqrt{1+\alphaD}}\, \label{Mtilde}
\ee
and
\be
\begin{split} 
\label{alphatilde}
\tilde{\alpha}_{\rm K} &=\frac{\alphaK+12\alphaB[\alphaC+(1+\alphaD)\alphaD]-6[ \alphaC+(1+\alphaD)\alphaD]^2 +3\Omega_{\rm m}\alphaD}{(1+\alphaC)^2(1+\alphaD)^2} \;, \\ \tilde{\alpha}_{\rm B}&= \frac{1+\alphaB}{(1+\alphaC)(1+\alphaD)}-1 \;,  \\
\tilde{\alpha}_{\rm M} &= \frac{\alphaM- 2 \alphaC }{1+\alphaC} - \frac{\dot \alpha _{\rm D}}{2 H \bar N (1+\alphaD) (1+\alphaC)} \;, \\
\tilde{\alpha}_{\rm T}&=(1+\alphaT)(1+\alphaD)-1\; . 
\end{split}
\ee
We can use these transformations, which depend on the two arbitrary functions $\alphaC$ and $\alphaD$, to set to zero any two of the parameters $\tilde\alpha_a$ above.

Finally, the conformal and disformal coefficients associated with the respective matter Jordan frame metrics are modified according to
\be
\label{alphaDC_change}
\begin{split}
\tilde{\alpha}_{\text{D},I} & = \frac{\alphaDI - \alphaD}{1+\alphaD} \;,  \\
\tilde{\alpha}_{\text{C},I}  &= \frac{\alphaCI - \alphaC}{1+\alphaC} \; .
\end{split}
\ee
Alternatively to setting two $\tilde\alpha_a$ to zero, it is always possible to choose as the new reference metric $\tilde g_{\mu\nu}$ one of the matter Jordan metrics, say 
$g_{\mu \nu}^{(I_*)}$, which then implies $\tilde{\alpha}_{\text{C},I_*}=\tilde{\alpha}_{\text{D},I_*}=0$.

One can verify that all the stability conditions are frame independent. In particular, the quantities that appear in the no-ghost conditions, eqs.~\eqref{alpha_def} and \eqref{NECdisf}, transform as
\be
\tilde \alpha= \frac{\alpha}{(1+\alphaC)^{2}(1+\alphaD)^{2}}  \, , \qquad \tilde \rho_I+(1+\tilde{\alpha}_{\text{D},I} )\tilde p_I= \frac{ \rho_I+(1+\alpha_{\text{D},I} ) p_I}{C^{2}(1+\alphaD)^{1/2}} \, ,
\ee
and since $1+\alphaD>0$ (see discussion in Sec.~\ref{sec:ma}), their sign is indeed frame independent.
It is also  straightforward to check that all the propagation speeds, i.e.~of tensor, scalar and matter fluctuations, transform in the same way and that their signs remain unchanged,
\be
\label{changecs}
 \tilde c_T^2 = (1+\alphaD)c_T^2\,,\quad \tilde c_s^2 = (1+\alphaD)c_s^2\,, \quad \tcI^2 =  (1+\alphaD)\cI^2\,.
 \ee

In summary, at the level of linear perturbations our gravitational sector is characterized by four time-dependent parameters $\alphaK$, $\alphaB$, $\alphaM$ and $\alphaT$. Each species is characterized by two time-dependent parameters, associated with their conformal and disformal couplings respectively. A priori, for a system of $N_S$ species coupled to different metrics, this gives  a total of $2N_S+4$ parameters. However, the general invariance of the system under an arbitrary change of frame, characterized by two parameters, reduces the number of independent parameters to $2(N_S+1)$. 

In particular, action \eqref{S2} can also be used to describe inflationary perturbations. In this case, matter can be ignored, i.e.~$N_S=0$, and one can always use eq.~\eqref{Mtilde} to find a frame where the Planck mass is time-independent and $c_T=1$, without loss of generality \cite{Creminelli:2014wna}. Thus, inflationary fluctuations can be generically described in the frame where $ \alpha_{\rm M} = 0 =  \alpha_{\rm T}$ by only {\it two} operators, those proportional to $\alphaK$ and $\alphaB$, as in Refs.~\cite{Creminelli:2006xe,Cheung:2007st}.\footnote{The sound speed of fluctuations in this case is 
$c_s^2 = - (2/\DDt)  \big[({1+\alphaB}) ( {\dot H}/{H^2}  +  \alphaB ) +  {\dot \alpha_{\rm B}}/{ H}  \big]$; see eq.~\eqref{ss}.
Thus, for a constant $\alphaB$, the usual gradient instability associated with the violation of the Null Energy Condition for $\dot H \geq 0$ can  be cured by requiring $ -1 \leq \alphaB \leq -\dot H/H^2$ \cite{Creminelli:2006xe}.}

\section{Matter equations of motion}
\label{sec3}

In this section, we leave  the unitary gauge description introduced in the previous section, by ``covariantizing'' the action. This can be done explicitly by performing a time reparametrization of the form
\be
t \to \phi = t+ \pi(t, \mathbf{x}) \;, \label{stuek}
\ee
where the unitary time $t$ becomes a four-dimensional scalar field $\phi$. For convenience, we denote by $\pi$ the fluctuation of $\phi$. 

By substituting the above transformation into the total action  $S = S_{\rm g} + S_{\rm m} $, we then obtain an action that depends on the scalar field $\phi$ and an arbitrary metric $g_{\mu\nu}$. We will use this more general form for the action to derive 
 the evolution equations for the gravitational and matter sectors.

The  equations of motion for the metric are obtained by varying the total action with respect to $g_{\mu\nu}$,
\be
\label{Einstein_eqs}
\frac{\delta S}{\delta g_{\mu \nu}} =0 \;. 
\ee
which provides the generalized Einstein equations. At linear order, they are explicitly given in Appendix~\ref{app1}.

To write the equations of motion for matter, we use the invariance of  the matter action $S_I$ under arbitrary diffeomorphisms, $x^\mu \to x^\mu + \xi^\mu$. This implies 
\be
\label{evol_matter}
\nabla_\mu T_{(I)}{}_{ \ \nu}^\mu + Q_I \partial_\nu \phi =0 \;, 
\ee
where the function $Q_I$, which characterizes the coupling between the matter species $I$ and the scalar field, is defined by
\be
\label{def_QI}
Q_I \equiv - \frac{1}{\sqrt{-g}} \frac{\delta S_I}{\delta \phi} = - \frac{ C_I' }{2 C_I} T_{(I)} - \frac{D_I'}{2C_I} T_{(I)}^{\mu \nu} \partial_\mu \phi \partial_\nu \phi + \nabla_\mu \left(T_{(I)}^{\mu \nu} \partial_\nu \phi \frac{D_I}{C_I} \right) \;,
\ee
where a prime denotes a derivative with respect to $\phi$.
The expression on the right hand side is obtained by using the property that the matter action $S_I$ depends on the scalar field only through the  Jordan metric eq.~\eqref{disf_unit_I2}.
 
Finally, the evolution equation for $\phi$ can be obtained by variation of the total action  with respect to $\phi$, $\delta S /\delta \phi =0$. Thus, from eq.~\eqref{def_QI} we obtain
\be
\label{pi_evol}
\frac{1}{\sqrt{-g}} \frac{\delta S_{\rm g}}{\delta \phi} -  \sum_I Q_I  =0 \;.
\ee

In the following, we will study the above equations, first in the homogeneous limit and then restricting ourselves to their  linearized version.

\subsection{Homogenous equations}
\label{sec:evol1}

Let us first consider the homogeneous case,  with the flat FLRW metric (\ref{FLRW}) where we set $\bar{N}=1$.
The  associated Friedmann equations are given in eqs.~\eqref{Fried1} and \eqref{Fried2}, or \eqref{Fr1} and \eqref{Fr2},  with $\rho_{\rm m} = \sum_I \rho_I$ and $p_{\rm m} = \sum_I p_I$.

For a FLRW background, the definition of $Q_I$, eq.~\eqref{def_QI},  reduces to
\be
\bar{Q}_I= \frac{H \rho_I}{1+\alphaDI}    \left\{     \alphaCI \left[ 1 - 3 w_I  (1+ \alphaDI) \right]   +   \alphaDI \left(3 +  \frac{\dot\rho_I}{H \rho_I}  \right)
+ \frac{\dot\alpha_{\text{D},I}}{2H (1+\alphaDI)}\right\}   \,, \label{barQ}
\ee
where we recall that the conformal and disformal parameters $\alphaCI$ and $\alphaDI$ are respectively defined in eq.~\eqref{alphaCD}. 
Substituting the above expression into eq.~\eqref{evol_matter}, one finds that the homogeneous matter evolution equation can be written in the form 
\be
\label{continuity_I}
 \dot \rho_I + 3 H(1+w_I -\gamma_I) \rho_I= 0 \;,
\ee
where the dimensionless parameter $\gamma_I$ is given by 
\be \label{gammab}
\gamma_I  \equiv  \frac13 \alphaCI \left[ 1- 3 w_I  (1+ \alphaDI) \right]   - w_I  \alphaDI  + \frac{\dot\alpha_{\text{D},I}}{6H(1+\alphaDI)}
 \,.
\ee
 Taking into account (\ref{continuity_I}), one can also check that  $\bQ_I = 3 H \rho_I \gamma_I $.
Note that the equation of state in the Jordan frame of the fluid $I$ corresponds to $\check w_I = w_I (1+ \alphaDI) $ (see Appendix \ref{app:bkg}). Using  this relation, one can check  that for a relativistic fluid, i.e.~$\check w_I = 1/3$, the conformal term in \eqref{gammab}
disappears, as expected from the tracelessness of its stress energy tensor.

Given the Friedmann equations \eqref{Fr1} and \eqref{Fr2} as well as the continuity equation for matter, eq.~\eqref{continuity_I}, the homogenous energy density of dark energy satisfies
\be
\dot \rho_{\rm DE} + \left[3 (1 + \wD) -  \alphaM \right] H \rhoD= H   \sum_I (\alphaM - 3 \gamma_I) \rho_I \;, \label{continuity_DE}
\ee
where we have introduced  the equation of state parameter for the dark energy component $
\wD   \equiv  {\pD}/{\rhoD}$.

\subsection{Perturbation equations in Newtonian gauge}

We now consider a linearly perturbed FLRW metric in Newtonian gauge with only scalar perturbations, i.e.,
\be
\label{metric_Newtonian}
ds^2 = - (1+2 \Phi) dt^2  + a^2(t) (1-2 \Psi) \delta_{ij}   dx^i dx^j \;.
\ee
In this gauge, we decompose the scalar part of the  stress-energy tensor for each species, at linear order,  as 
\begin{align}
T_{(I)}{}^0_{\   0} &\equiv -  (\rho_I+ \delta \rho_I) \;, \label{se1}\\
T_{(I)}{}^0_{\   i} & \equiv \rho_I(1 + w_I)  \partial_i  v_I = - a^2 T_{(I)}{}^i_{\ 0}\;, \label{se2} \\
T_{(I)}{}^i_{\   j} &\equiv (\rho_I w_I  +  \delta p_I) \delta^i_j + \left( \partial^i \partial_j - \frac13 \delta^i_j \partial^2 \right) \sigma_I \label{se3}\;,
\end{align}
where  $\delta \rho_I $ and $\delta p_I$ are the  energy density and pressure perturbations, $v_I$ is the 3-velocity potential and $\sigma_I$ is the  
anisotropic stress potential for the species $I$. As usual, we define the total matter quantities as $\delta \rho_{\rm m} = \sum_I \delta \rho_I$, $\delta p_{\rm m} = \sum_I \delta p_I$, $v_{\rm m} = \sum_I (\rho_I + p_I) v_I/(\rho_{\rm m} + p_{\rm m}) $ and $\sigma_{\rm m}= \sum_I \sigma_I$.

The  continuity equation, for each species,  can be derived from the time component of eq.~\eqref{evol_matter}.
In Fourier space, at linear order, this reads
\be
\delta \dot \rho_I + 3 H (\delta \rho_I + \delta p_I)- 3 \rho_I (1+w_I) \dot \Psi - \rho_I(1 + w_I) \frac{k^2}{a^2} v_I =  \bar Q_I \dot \pi + \delta Q_I \;,
\ee
where $\bar Q_I$ and $\delta Q_I$ are given respectively by eqs.~\eqref{barQ} and \eqref{Qpert}.
The space components of  eq.~\eqref{evol_matter} gives the Euler equation, which at linear order reads 
\be
\rho_I(1 +w_I )\dot{v}_I+  \rho_I\left[\dot w_I-3Hw_I(1+w_I)\right]  v_I +  \delta p_I+ \rho_I(1+w_I)\Phi - \frac{2 }{3 } \frac{k^2}{a^2}\sigma_I= -
\bar Q_I \big[  \pi  +v_I(1+w_I) \big]\;.
\ee

We can rewrite the equations above in terms of the density contrast  $\delta_I \equiv \delta \rho_I /\rho_I$
and using the explicit expression for $\delta Q_I$ given in eq.~\eqref{Qpert}. This yields
\be
\label{continuity_pert}
\begin{split}
& \dot \delta_I + 3H (1 + \alphaCI)(1 + \alphaDI) \left(\frac{\delta p_I}{\rho_I} - w_I\delta_I\right) - (1+w_I) \frac{k^2}{a^2} v_I =  3 \left[1+ (1+\alphaDI)w_I\right] \dot \Psi 
\\ 
& 
+ 2(1+\alphaDI) \left[ \alphaCI (1-3 w_I) - 3 \gamma_I\right] H \Phi -\alphaDI \left( \dot \Phi - \ddot \pi + w_I \frac{k^2}{a^2} {\pi} \right) \\
& -  \left[2 (1+\alphaDI)\alphaCI (1 - 3 w_I) - 3 w_I \alphaDI - 3 \gamma_I (3+ 2\alphaDI )\right]H  \dot \pi \\
& + 3 \left[ \left( \gamma_I H\right)^{\hbox{$\cdot$}} + w_I \alphaDI \dot H +\left(\alphaCI + \alphaDI(1 + \alphaCI)\right) \dot w_I  H \right] {\pi} \;,
\end{split}
\ee
and
\be \label{euler_pert}
\dot{v}_I- 3 H \left[ w_I - \gamma_I - \frac{\dot w_I }{ 3H(1+w_I)} \right] v_I +  \frac{\delta p_I}{(1+w_I)\rho_I} + \Phi - \frac{2 }{3 (1+w_I) \rho_I} \frac{k^2}{a^2}\sigma_I= - 3 H \frac{ \gamma_I}{1 + w_I} \pi \;.
\ee
As mentioned before, the equation of state parameter in the  matter Jordan frame, $\check w_I$, is different from the one in a generic frame, $w_I$. This means that the relation between pressure  and energy density perturbations  depends on the frame. Indeed, because of the coupling  to the scalar field, there is a non-adiabatic pressure perturbation \cite{Malik:2004tf} which appears in frames that are disformally distinct from the Jordan one (see also \cite{Minamitsuji:2014waa} for a similar remark).
For an isentropic perfect fluid with  $\check c^2_{s,I} = \check w_I =$ constant,~this reads  (see Appendix \ref{sec:B2})
\be
\delta p_{\text{nad}, I} \equiv {\delta p_I} - \frac{\dot p_I}{ \dot \rho_I}   \delta \rho_I =  p _I  \left[2  \alphaD(\Phi-\dot \pi) + \frac{\dot \alpha_{\rm D}}{1+\alphaD} \bigg( \frac{\delta \rho_I}{\dot \rho_I}  - \pi \bigg)\right]\, . \label{eqs_pert}
\ee

Let us comment on the initial conditions of the above equations. In the simplest case, one can assume that perturbations start in the  adiabatic growing solution, which is justified if they have originated from single-field inflation (see e.g.~\cite{Weinberg:2003sw}). In this case, their  amplitude can be given in terms of the time-independent quantity ${\cal R}_{\rm in}$,  defined as the long-wavelength limit ($k \ll aH$) of the total comoving curvature perturbation \cite{Bardeen:1983qw}
\be
{\cal R} \equiv  - \Psi + H\frac{ \dot \Psi + H\Phi}{\dot H} \;.
\ee
In Ref.~\cite{Gleyzes:2014rba} it was shown that, in the absence of nonminimal couplings, the generalized Einstein equations and the evolution equations for the matter and field fluctuations admit the adiabatic solution 
\be
\label{adia_sol}
\begin{split}
\Phi & = - (1+\alphaT) {\cal R}_{\rm in}  + (1+\alphaM)  H \epsilon - \frac{\sigma_{\rm m}}{M^2} \;, \qquad
\Psi  =  - {\cal R}_{\rm in} + H \epsilon \;, \\
\delta \rho_I &= -  \dot \rho_I \epsilon   \;, \qquad \delta p_I = -  \dot p_I \epsilon   \;, \qquad
v_I  =  \epsilon \;, \qquad
\pi  = -\epsilon \;,
\end{split}
\ee
where 
\be
\epsilon \equiv \frac{1}{M^2 a} \int a \left[ M^2 (1+\alphaT) {\cal R}_{\rm in} + \sigma_{\rm m}  \right] dt  \;. \label{epsilon_def}
\ee
One can check that these expressions are frame invariant and remain a solution even in the presence of nonminimal couplings, with the same ${\cal R}_{\rm in}$. Note that, for adiabatic initial conditions, the right hand side of \eqref{eqs_pert} automatically vanishes and that the matter perturbations are in effect adiabatic in all frames. The nonadiabatic pressure term due to the change of frame manifests itself only for nonadiabatic initial conditions.

Let us point out that the equations written in this section include as a special case (corresponding to $\alphaM=\alphaT=\alphaB=0$ and $\alphaDI=0$) the equations of motion for  linear perturbations derived in standard models of dark energy ($k$-essence) conformally coupled with matter (see e.g. \cite{Amendola:2003wa}). Our results also include the more recent investigations of disformal couplings between matter, usually CDM, and some standard dark energy (i.e.~with $\alphaM=\alphaT=\alphaB=0$) \cite{vandeBruck:2012vq,Koivisto:2012za,Zumalacarregui:2012us,vandeBruck:2015ida,Minamitsuji:2014waa}. 

In the general case, eqs.~\eqref{continuity_pert}--\eqref{euler_pert}  can be directly applied to the usual matter species, i.e.~CDM, baryons, photons and neutrinos and implemented in a numerical code. If one wants to study the CMB fluctuations, the fluid approximation is not sufficient for photons and neutrinos and must be replaced by a Boltzmann description. Whereas a nonminimal coupling of photons is constrained to remain tiny \cite{Dupays:2006dp}, one could envisage a nonminimal coupling of neutrinos (see e.g.~\cite{Fardon:2003eh,Brookfield:2005td,Afshordi:2005ym}). To deal with this modification, the simplest method would consist in writing the Boltzmann equation in the Jordan frame of the neutrinos, where it keeps its usual form. The neutrino-frame gravitational potentials appearing in this equation could then be expressed in terms of  the gravitational potentials $\Phi$ and $\Psi$ associated with the baryon-photon frame, by using explicitly the disformal transformation between the two frames, as given in Appendix~\ref{app2}.

\section{Baryons and coupled CDM}
\label{sec4}

In this section, we apply the general formalism developed in the previous sections to the cosmological era where the dominant matter species are baryons (denoted by the subscript $b$) and CDM (subscript $c$). Whereas there exist very stringent constraints on EP violation for baryons \cite{Will:2014xja,Brax:2014vva}, the dark matter sector is much less constrained \cite{Ade:2015rim}. For this reason, we now assume that the baryons are minimally coupled, i.e.
\beq
\alphaCb=0\,, \qquad \alphaDb=0\qquad \Rightarrow \qquad \gamma_b=0\,,
\eeq
  while dark matter is coupled to dark energy via a general metric of the form~\eqref{disf_unit_I}.

For both baryons and CDM, one neglects the pressure and anisotropic stress, so that   $w_{b}=w_{c}=0$ for the background and $\delta p_b=\delta p_c=\sigma_b=\sigma_c=0$ for the perturbations~\footnote{As shown  in Appendix~\ref{sec:B2}, this statement holds in any frame.}. 

The background equations \eqref{continuity_I} and \eqref{continuity_DE} take the form
\begin{align}
\dot \rho_b + 3 H \rho_b &= 0 \;, \label{cons_1}\\
\dot \rho_c + 3 H (1- \gamma_c) \rho_c & = 0\;, \label{cons_2} \\
\dot \rho_{\rm DE}+  \left[ 3 (1 +\wD) -  \alphaM \right]  H \rhoD  &=  -3 H \gammac \rho_c +H\alphaM \rho_{\rm m}  \label{cons_3} \;.
\end{align}
According to (\ref{gammab}), the coupling parameter $\gammac$ is related to the CDM  conformal and disformal parameters via 
\be
\gammac=\frac13 \alphaCc+\frac{\dot \alpha_{\text{D},c}}{6H(1+\alphaDc)}\,.
\ee

\subsection{Linear perturbations}
Let us now consider the linear perturbations. The continuity and Euler equations, (\ref{continuity_pert}) and (\ref{euler_pert}), reduce to  
\begin{align}
\dot \delta_b  - \frac{k^2}{a^2}v_b  & =  3\dot \Psi   \, , \label{cont_b} \\
 \dot v_b  & = -  \Phi  \, , \label{eul_b} \\
 \dot \delta_c -\frac{k^2}{a^2}v_c    & = 3 ( \Psi + \gammac H \pi )^{\hbox{$\cdot$}} + 
2({1+\alphaDc})({\alphaCc- 3\gammac }) H (\Phi - \dot \pi)-\alphaDc (\dot \Phi-\ddot \pi )  \label{cont_c} \, , \\
\dot v_c +3H\gammac v_c & =- \Phi  - 3H\gammac \pi \, . \label{eul_c}
\end{align}
These equations must be supplemented by the  Einstein equations, eqs.~\eqref{Ein1}--\eqref{Ein4} and by the scalar fluctuation equation \eqref{sfee}.

It is possible to use a combination of the Einstein equations and of \eqref{sfee} to eliminate the dependence on $\pi$ and $\dot \pi$ in the above equations in favour of the gravitational potentials. The same procedure has been used in the case of minimally coupled matter in Refs. \cite{Bellini:2014fua} and \cite{Gleyzes:2014rba}. In our baryon and coupled CDM system we find
a dynamical equation for $\Psi$ of the form:
\beq
\begin{split}
&\ddot \Psi +  \frac{\mbeta_1 \nbeta_2 + \nbeta_3  \alphaB^2 \,\kH^2}{\mbeta_1+\alphaB^2 \kH^2} H \dot \Psi+   \frac{ \mbeta_1 \nbeta_4+\mbeta_1 \nbeta_5  \,\kH^2   +c_s^2  \alphaB^2 \kH^4} {\mbeta_1+ \alphaB^2 \,\kH^2 }  H^2  \Psi = \\
&- \sum_{I}\frac{3 }{2 } H^2 \Omega_I \Bigg[  \frac{\mbeta_1\nbeta_{6,I} + \nbeta_{7,I}  \alphaB^2 \, \kH^2  }{\mbeta_1+ \alphaB^2 \,\kH^2  }   \delta_{I}  + \frac{\mbeta_1 \nbeta_{8,I} + \nbeta_{9,I}  \alphaB^2\, \kH^2  }{\mbeta_1+ \alphaB^2 \,\kH^2 } H  v_{I}   \Bigg]\; , \label{Phi_evol}
\end{split}
\eeq
where $\kH \equiv k/(aH)$ and the time-dependent coefficients   $\nbeta_a$ are explicitly given  in Appendix~\ref{parameters}. They generally differ  from those given in Refs.~\cite{Bellini:2014fua} and\cite {Gleyzes:2014rba} because the  disformal coupling to dark matter modifies the evolution equation for $\pi$, see eq.~\eqref{sfee}.
The relation between  $\Phi$ and $\Psi$ is given by
\beq
\label{Phitopsi}
\begin{split}
&\alphaB^2 \kH^2 \left( \Phi - \Psi \frac{\xi}{\alphaB}   \right)  + \mbeta_1  \left[ \Phi - \Psi ( 1 + \alphaT )\left(1+\DD \frac{\alphaT-\alphaM }{2\beta_1}\right)
 \right] = \\
&\frac {\alphaT-\alphaM }{ 2}\left\{  \DD \frac{ \dot \Psi}{H}+3\sum_I\Omega_I  \left[ \alphaB\delta_I+\frac{\alphaK-6\alphaB}2H v_I  \right]\right\} \;,
\end{split}
\eeq
where $\DD\equiv \alphaK + 6\alphaB^2$ and we have introduced  the time-dependent combination
\be
\xi \equiv   \alpha_B (1+\alpha_T) + \alpha_T - \alpha_M \;. \label{xi_def}
\ee
For $\alphaT=\alphaM=0$, this  reduces to the familiar relation $\Phi=\Psi$. 

We can also eliminate the dependence on $\pi$ from the continuity and Euler equations for CDM, eqs.~\eqref{cont_c} and \eqref{eul_c}.
For simplicity, we give here the explicit expressions only in the case $\alphaM=\alphaT=0$, for which $\Phi=\Psi$, the generalization being straightforward. 
In this case, the continuity and Euler equations \eqref{cont_c} and \eqref{eul_c} become
\begin{align}
\dot \delta_c-\frac{k^2}{a^2}v_c=&\, \frac{\mbeta_1 \xi_2 + \xi_3  \,\kH^2}{\mbeta_1+\alphaB^2 \kH^2}  \dot \Psi+   \frac{ \mbeta_1 \xi_4+\mbeta_1 \xi_5  \,\kH^2   +c_s^2  \alphaB \frac{\alpha_{\text{D},c}}{1+\alpha_{\text{D},c}} \kH^4} {\mbeta_1+ \alphaB^2 \, \kH^2 }  H  \Psi \nonumber \\
& +\sum_{I}\frac{ 3}{2} \Omega_{I} H \bigg[  \frac{\mbeta_1\xi_{6,I} + \xi_{7,I}  \, \tilde k^2  }{\mbeta_1+ \alphaB^2 \, \kH^2  }   \delta_{I}  + \frac{\mbeta_1 \xi_{8,I} + \xi_{9,I}  \, \kH^2  }{\mbeta_1+ \alphaB^2 \, \kH^2 } H  v_{I}   \bigg]\;, \label{conteqc2} \\
\dot v_c+ 3H\gamma_c v_c=&\, -\Psi- \frac{3\gamma_c}{\mbeta_1+\alphaB^2 \kH^2}\bigg\{\DD\frac{\dot \Psi +H\Psi}{2H}+\kH^2 \alphaB \Psi  \nonumber \\
&+\sum_I \frac{ 3}{2} \Omega_{I} H \left[\alphaB\delta_I+\frac{\alphaK- 6\alphaB\gamma_I}2H v_I\right]\bigg\}\,, \label{Eulerc2}
\end{align}
where the time-dependent coefficients $\xi_a$ are given in Appendix~\ref{parameters}. In this case, where $\Phi=\Psi$, eqs.~\eqref{cont_b}, \eqref{eul_b}, \eqref{Phi_evol}, \eqref{conteqc2} and \eqref{Eulerc2} form a closed system of equations.

\subsection{Quasi-static approximation}
\label{QSCDM}
To investigate {late-time} cosmology, it is convenient to resort, on sufficiently short scales, to the quasi-static limit. This is justified as long as we remain on scales smaller than the sound horizon of dark energy, i.e.~$k \gg aH/c_s$ \cite{Sawicki:2015zya}. In this limit, the conservation and Euler equations for baryons and CDM (eqs.~\eqref{cont_b}--\eqref{eul_c}) simplify {to}
\begin{align}
\dot \delta_b  - \frac{k^2}{a^2}v_b  & =  0  \, , \label{conteqb}\\
 \dot v_b  & = -  \Phi  \, , \label{Eulerb} \\
\dot \delta_c -\frac{k^2}{a^2}v_c    & = 0   \, ,  \label{conteqc}\\
\dot v_c +3H\gammac v_c & =- \Phi  - 3H\gammac \pi \, .\label{Eulerc}
\end{align}
In these equations, all the modifications are encoded in the {single} parameter $\gamma_c$. Therefore, it is not possible to disentangle the conformal and disformal effects. Note that this is due to the fact that the nonminimally coupled species is pressureless and that we restrict to the quasi-static regime.

We can then use the generalized Einstein equations to derive the Poisson equation for $\Phi$. Combining eqs.~\eqref{Ein1} and \eqref{Ein3} one finds
\be
\label{poisson}
- \frac{k^2}{a^2} \Phi  =  \frac{3}{2} H^2 \Omega_{\rm m}\left\{ \left(1+\alphaT+ \tbeta_\xi^2 \right) \omega_b \delta_b +  \left[ 1+\alphaT+ \tbeta_\xi ( \tbeta_\xi  +  \tbeta_\gamma) \right] \omega_c \delta_c \right\}  \,, \quad \omega_I\equiv\frac{\Omega_I}{\Omega_{\rm m}}\, ,
\ee
where we have defined the dimensionless parameter
\be
\label{pi}
\tbeta_{\gamma}  \equiv \frac{3 \sqrt{2}  \gamma_c }{ c_s \DDt^{1/2}  } \; ,
\ee
which characterizes the strength of the nonminimal coupling of CDM, 
as well as the analogous parameter
\be
\tbeta_{\xi} \equiv \frac{\sqrt{2} \xi }{  c_s \DDt^{1/2} } \;,  \label{betaxi_def}\\
\ee
associated with the  modified gravity coefficient $\xi  $ defined in \eqref{xi_def}.
Note that the denominator in the definitions of $\tbeta_{\gamma}$ and $\tbeta_{\xi}$ is real, since stability 
 requires that $c_s^2  \DDt \ge 0$. 

Moreover, in the quasi-static limit the evolution equation \eqref{sfee} for $\pi$ reduces to a constraint equation, which reads
\be
\label{pipoisson}
 - \frac{k^2}{a^2} \pi = 3H \Omega_{\rm m}\frac{ \tbeta_\xi \omega_b \delta_b +  (\tbeta_\xi + \tbeta_\gamma)  \omega_c \delta_c }{\sqrt{2} c_s \DDt^{1/2} }  \;.
\ee
Substituting (\ref{poisson}) and (\ref{pipoisson}) into the matter equations~\eqref{conteqb}--\eqref{Eulerc}, we obtain
two coupled second-order differential equations for the two density contrasts:
\begin{align}
\ddot \delta_b + 2 H \dot \delta_b &= \frac32  H^2  \Omega_{\rm m} \left\{ (1+\alphaT+ \tbeta_\xi^2 ) \omega_b \delta_b +  \left[ 1+\alphaT+ \tbeta_\xi ( \tbeta_\xi  +  \tbeta_\gamma) \right] \omega_c \delta_c \right\}  \;, \label{eqmatterb}\\
\ddot \delta_c + (2 - 3 \gammac) H \dot \delta_c &= \frac32  H^2 \Omega_{\rm m}  \left\{  \left[ 1+\alphaT+ \tbeta_\xi ( \tbeta_\xi  +  \tbeta_\gamma) \right] \omega_b \delta_b +  \left[ 1+\alphaT+  ( \tbeta_\xi + \tbeta_\gamma )^2 \right]   \omega_c \delta_c  \right\}   \;.\label{eqmatterc}
\end{align}
In the absence of nonminimal coupling of CDM ($\beta_\gamma=0$), the gravitational coupling of both species is modified by the same factor $1+ \alphaT + \beta_\xi^2$.
In the absence of modified gravity ($\tbeta_\xi=0$ and $\alpha_T=0$), one finds that the nonminimal   coupling of CDM  ($\tbeta_\gamma\neq 0$)  modifies the friction term for $\delta_c$, as well as  increases  the  coefficient in front of $\delta_c$ in the second equation, whereas all other three coefficients on the right hand sides are unchanged. This is the result obtained in the context of coupled dark energy (see e.g. \cite{Amendola:2003wa}).
  By contrast, if one combines modified gravity ($\tbeta_\xi\neq 0$) with a nonminimal coupling of CDM, all four coefficients on the right hand sides are modified. We leave for the future the detailed study of how these new coefficients parametrize the influence of modified gravity on structure formation.

Let us now turn to  the two gravitational potentials $\Phi$ and $\Psi$.  
When considering the impact of dark energy on observations, it is often convenient to express the new relations between the two potentials $\Psi$ and $\Phi$ and the total matter density fluctuations in terms of modifications of the Newton constant. We thus introduce the parameters 
\be
\mu_\Phi \equiv - \frac{ 2 M^2 k^2 \Phi}{a^2 \rho_{\rm m} \delta_{\rm m}}  \;, \qquad \mu_\Psi \equiv - \frac{ 2 M^2 k^2 \Psi}{a^2 \rho_{\rm m} \delta_{\rm m}}   \;,
\ee
which are equal to one in the standard case. From eq.~\eqref{poisson} and an analogous Poisson-like equation for $\Psi$, obtained by combining the Einstein equations, one finds that the above parameters are given by
\begin{align}
\mu_\Phi  &= 1+\alphaT +   \tbeta_\xi \big( \tbeta_\xi  +  \tbeta_\gamma  \omega_c b_c  \big)\;, \label{muphi}\\
\mu_\Psi  &= 1+  \tbeta_{\rm B}  \big(  \tbeta_\xi +    \tbeta_\gamma  \omega_c b_c  \big) \;,  \label{mupsi}
\end{align}
where  we have defined 
\be
\tbeta_{\rm B}  \equiv \frac{\sqrt{2}  \alphaB }{ c_s \DDt^{1/2} } \;.
\ee
We have also introduced a time-dependent bias parameter, $b_c \equiv \delta_c/\delta_{\rm m}$.\footnote{In the quasi-static limit, the evolution equations \eqref{eqmatterb} and \eqref{eqmatterc} are scale independent so that the ratios $\delta_b/\delta_{\rm m}$ and $\delta_c/\delta_{\rm m}$ do not depend on scales.}

As the gravitational lensing effect depends on the sum of the two potentials, the relevant quantity parametrizing deviations in weak lensing observables (and equal to two in the standard case) is 
\be
\mu_{\rm WL} = \mu_\Psi +\mu_\Phi = 2+\alphaT +  (\tbeta_{\rm B}  + \tbeta_\xi ) \big(  \tbeta_\xi +    \tbeta_\gamma  \omega_c b_c  \big).
\ee
Thus, the impact of modifications of gravity due to  non-vanishing $\alphaB$, $\alphaM$ and $\alphaT$ affects   observable quantities in the perturbations through  $\alphaT$ and the combinations $\tbeta_{\rm B}$, $\tbeta_\xi$. Analogously, the effect of nonminimal couplings on observations is parameterized by $\tbeta_\gamma$ only (see the next section for the case of a coupled relativistc fluid, where another quantity is needed to parameterize the nonminimal coupling).
Note that as a consequence of dropping time derivatives in the fluctuations of $\pi$, the parameter $\DDt$ always appears  multiplied by $c_s^2$. From the definition of the sound speed, eq.~\eqref{ss}, $c_s^2 \DDt$ is independent of $\alphaK$,  so that the latter cannot be constrained by observations in the quasi-static limit \cite{Piazza:2013pua}.

When $\tbeta_\gamma=0$, i.e.~if CDM is minimally coupled and there are no EP violations, the last term inside the parenthesis of {eqs.~\eqref{muphi} and \eqref{mupsi}} drops  and these relations simplify  to 
\be
\mu_\Psi  =  1+ \tbeta_\xi \tbeta_{\rm B}  \;, \qquad \mu_\Phi  =  1+\alphaT+ \tbeta_\xi^2 \;.
\ee
In this case, the so-called slip parameter becomes (see for instance \cite{Gleyzes:2014rba})
 \be
\frac{\Psi}{\Phi} = \frac{1+ \tbeta_\xi \tbeta_{\rm B} }{1+\alphaT+ \tbeta_\xi^2}\,.
\ee 
By contrast, if there is a non trivial coupling of CDM but  gravity itself is not modified, in which case we have $\alphaB=0$, $\alphaM=0$ and $\alphaT=0$ (thus $\xi=0$), we recover that the Newton constant is not modified, $\mu_\Phi = \mu_\Psi =1$, and that $\Phi$ and $\Psi$ are the same as in GR, {\em even if} CDM is nonminimally coupled, as is the case in usual scenarios of coupled dark energy. In general, we find that the situation is much richer when both gravity and matter couplings are modified.

\section{Matter and coupled relativistic fluid}

\label{sec5}
\newcommand\alphaDnu{\alpha_{\text{D}, \text{r}}}
\newcommand\alphaCnu{\alpha_{\text{C},\text{r}}}

In this section, as another example we consider nonminimally coupled relativistic particles, in the fluid approximation. They could represent neutrinos, radiation or warm dark matter in the relativistic regime. Baryons and CDM are taken to be minimally coupled, $\alpha_{\rm D ,m}=\alpha_{\rm C, m}= 0$. 
In the Jordan frame of the relativistic fluid, its equation of state parameter is given by $\check w_{\text{r}} = 1/3$ (and $\check c_{s,\text{r}} ^2= 1/3$). Thus, in the frame where baryons and CDM are minimally coupled, the background and perturbed equations of state are
\be
\label{changepr}
w_\text{r} = \frac{1}{3(1+\alphaDnu)} \;, \qquad \delta p_{\text{r}}=\frac{\rho_{\text{r}}}{3(1+\alphaDnu)}\left[ \delta_{\text{r}} + 2\alphaDnu(\Phi-\dot\pi)-\frac{\dot \alpha_{\text{D},\text{r}}}{1+\alphaDnu}\pi \right]\, .
\ee
The second relation has been obtained from eq.~\eqref{eqs_pert}, using eq.~\eqref{changecs} for the sound speed. To simplify the treatment, we ignore the anisotropic stress, i.e.~$\sigma_{\text{r}} =0$.

We are now going to assume that baryons and CDM dominate the gravitational perturbations, thus neglecting the backreaction of the relativistic fluid. On small scales, we can then resort to the quasi-static approximation.
Under these conditions, 
the evolution equations for matter are
\be
\dot \delta_{\rm m}  - \frac{k^2}{a^2}v_{\rm m}   =  0   \, ,  \qquad
 \dot v_{\rm m}   = -  \Phi  \, . \label{eul_m_cont_m} 
 \ee
For the relativistic fluid, we use eqs.~\eqref{continuity_pert} and \eqref{euler_pert}
and replace $\gamma_{\text{r}}$ with the expression $\gamma_{\text{r}} =(\dot \alpha_{\text{D},\text{r}}-2H\alphaDnu)/[6H(1+\alphaDnu)] $.
The evolution equations then read
\begin{align} 
\dot \delta_{\text{r}}  -  \frac{4+3\alphaDnu}{3(1+\alphaDnu)}  \frac{k^2}{a^2} v_{\text{r}}   =&   -\frac{\alphaDnu }{3(1+\alphaDnu)} \frac{k^2}{a^2} \pi  \;, \\
  \dot v_{\text{r}} -H \left[1-3\, \frac{2+3\alphaDnu}{1+3\alphaDnu}\left(g_{\text{D}, {\text{r}}}-f_{\text{D}, {\text{r}}}\right)\right]v_{\text{r}} +\frac{\delta_{\text{r}}}{4+3\alphaDnu}  = &- (1 + 2f_{\text{D}, {\text{r}}} ) \Phi + 2f_{\text{D}, {\text{r}}}  \dot \pi -3Hg_{\text{D}, {\text{r}}} \pi\, , \label{eul_nu}
 \end{align}
where we have defined
\be
f_{\text{D},\text{r}}\equiv -\frac{\alphaDnu}{4+3\alphaDnu} \, , \qquad g_{\text{D}, {\text{r}}}  \equiv f_{\text{D},\text{r}} \left[1- \frac{(1+3\alphaDnu)\dot \alpha_{\text{D},\text{r}}}{6(1+\alphaDnu)H\alphaDnu}\right] \;.
\ee
As expected,  for $\alphaDnu=0$ the effects due to the nonminimal coupling vanish. Note that, even if we are in the quasi-static limit, the term in $\dot \pi$ should be kept in the Euler equation, as it is expected to be of the order of $H\pi$ and therefore comparable to the other terms.

Our assumption that the relativistic fluid does not contribute to the gravitational perturbations means that $\Phi$ and $\pi$ are only sourced by CDM and baryons, i.e.
\begin{align}
\label{poissonNu}
- \frac{k^2}{a^2} \Phi  &=  \frac{3}{2} H^2 \Omega_{\rm m}   \left(1+\alphaT+ \tbeta_\xi^2 \right)  \delta_{\rm m} \, ,\qquad - \frac{k^2}{a^2} \pi = 3H\Omega_{\rm m} \frac{ \tbeta_\xi   }{\sqrt{2} c_s \DDt^{1/2} } \delta_{\rm m}   \;,
\end{align}
 which correspond to eqs.~\eqref{poisson} and \eqref{pipoisson} specialized to the case $\beta_\gamma=0$. Therefore, even when the extra scalar field is not sourced by matter perturbation (e.g. when $\beta_{\xi}=0$) and $\pi=0$, the relativistic particles still feel a force $F_{\rm r}$ different from that felt by matter, $F_{\rm m}$, the relative difference being given by $(F_{\text r}-F_{\text m})/F_{\text m}= 2f_{\text{D}, {\text{r}}}$. This extra force is due to the non-adiabatic pressure perturbation $\delta p_{\text{nad}, I}$ in eq.~\eqref{eqs_pert}, induced by the disformal coupling out of the Jordan frame of the fluid (see eq.~\eqref{changepr}). 
 
 To highlight this effect, the Euler and continuity equations can be combined to form a second-order differential equation for the density contrast $\delta_{ {\text{r}}}$, sourced by the matter perturbations according to eq.~\eqref{poissonNu}. In the simple case where gravity is not modified, i.e.~$\beta_{\xi}=\alpha_T=0$, we get
\be
\begin{split}
&\ddot{\delta}_{ {\text{r}}}+ H \dot{\delta}_{ {\text{r}}} \left(1+3\alphaDnu \frac{1- g_{\text{D},\text{r}}/ f_{\text{D},\text{r}}}{1+3\alphaDnu} \right)+\frac{k^2}{3 a^2(1+\alphaDnu)}\delta_{ {\text{r}}}=2  H^2\Omega_{\rm m}  \frac{ 1+ \alphaDnu/4 }{1+\alphaDnu} \, \delta_{\rm m} \, .
\end{split}
\ee
 Unlike in the case of the bayons-CDM fluid, the signature of the disformal coupling here is present at the linear level in $\alphaDnu$, while in eqs.~\eqref{eqmatterb}--\eqref{eqmatterc} it appears at the quadratic level through the terms $ \beta_{\xi} \beta_{\gamma} $ and $ \beta_{\gamma}^2$.

The main message of this section is that one must define the usual fluid properties (such as the equation of state and the speed of sound) in the Jordan frame, where the species is minimally coupled to gravity.

\section{Conclusions}
\label{sec_last}

In this work, we have presented an effective description of dark energy and modified gravity, which extends the approach developed in \cite{Gleyzes:2013ooa} by relaxing the assumption of  universal coupling of all matter species. Namely, we have allowed each matter species to be associated with a specific Jordan frame (or metric), conformally and disformally related to the gravitational metric. In this way, we have made connection with a vast sector of the literature devoted to the so-called coupled dark energy, with either a conformal coupling in most works or a disformal coupling for more recent works. 
However, in contrast with this previous  literature, we have considered here a very general description of the gravitational sector, which includes Horndeski's theories (although not their extensions such as G$^3$) {instead of} general relativity with a quintessence-like scalar field as usually assumed.

At the level of linear perturbations, the gravitational sector is described by the quadratic action given in eq.~\eqref{S2}, which depends on four time-dependent parameters $\alphaK$, $\alphaB$, $\alphaM$ and $\alphaT$. As for matter, each species is characterized by two time-dependent parameters, $\alphaCI$ and $\alphaDI$, associated with their conformal and disformal couplings to the gravitational metric. This implies that the whole system depends on a total of $2N_S+4$ time-dependent parameters,  if  $N_S$ species are present.
However, there is some arbitrariness in the choice of the gravitational metric that is used to define the gravitational and matter sectors. By considering a conformal-disformal transformation \eqref{disf_uni} of this metric, the same physical system is characterized by $2N_S+4$ new parameters, 
which transform according to \eqref{alphatilde} and \eqref{alphaDC_change}. Taking into account this ``gauge'' redundance, which depends on two arbitrary parameters, one thus finds that the number of physically relevant parameters is reduced to $2(N_S+1)$. 

A very useful result of the present work is the derivation of the linear stability conditions in this very general framework. As the presence of disformal couplings contributes to the kinetic energy of the scalar fluctuations, the  condition for the absence of ghosts is modified. This now requires that $\alpha$ defined in eq.~\eqref{alpha_def} is positive. We have checked that the stability conditions are invariant under the ``gauge transformations'' of the parameters discussed above. 

We have also written the equations of motion for the linear perturbations and emphasized how the usual equations are modified in the presence of modified gravity and nonmininal (conformal or disformal) couplings. Special care must be taken when the chosen frame does not coincide with the matter Jordan frame as the  relations between matter quantities are frame-dependent. For instance, the equation of state parameter, whose natural value (e.g. $1/3$ for radiation) is defined in the Jordan frame associated with the matter species, will be in general different in another frame. 

We have illustrated our formalism by considering two types of scenarios, motivated by the  already stringent constraints on the nonminimal coupling of ordinary species (baryons and photons) to a scalar field. In the first case, we have focused our attention to the situation where only CDM is nonminimally coupled to the scalar field. For late cosmology, in the quasi-static approximation, we have computed the evolution equations of CDM and baryon density contrasts. In the second case, we have assumed that both baryons and CDM are minimally coupled but allowing for a relativistic fluid (e.g.~neutrinos) with nonminimal couplings. These two simple examples illustrate what kind of new effects can be produced by the combination of modified {gravity} and nonminimal couplings. 

It would  be interesting to investigate how future observations will be able to constrain simultaneously the parameters describing the deviations from GR and those characterizing the coupling of matter to this generalized gravitational sector.

\vspace{5mm}
\noindent
{\bf Acknowledgements}.
We thank Federico Piazza and Ignacy Sawicki for stimulating discussions. DL and FV acknowledge the participation to the workshop ``The Quest for Dark Energy II'' at Ringberg Castle, where part of this work was done. JG, MM and FV thank the APC ({\em AstroParticule et Cosmologie}) and PCCP (Paris Center for Cosmological Physics) for kind hospitality. FV acknowledges partial support from the grant ANR-12-BS05-0002 of the French Agence Nationale de la Recherche. 

\appendix

\section{Changing frame}
\label{app2}

We consider a general disformal transformation of the metric \eqref{disf_trans}, which in unitary gauge reads 
\be
\label{disf_uni_App}
g_{\mu \nu} \to \tilde g_{\mu \nu} = C(t) g_{\mu \nu}  + D(t) \delta_\mu^0 \delta_\nu^0  \;,
\ee
and study how metric and matter quantities change under this transformation.
In terms of  the two  time-dependent parameters $ C$ and $ D$,
the  ADM components of the new metric $\tilde g_{\mu \nu}$ in unitary gauge are given by
\be 
\label{disformalADM}
{\tilde{N}}^2={C}N^2 - D \;, \qquad \tilde{N^i}=N^i \, ,\qquad \tilde{h}_{ij}=C h_{ij}   \, ,
\ee
while the intrinsic Ricci scalar and the extrinsic curvature respectively transform as
\be \label{disformalRK}
\tilde {R}=C^{-1} R \, , \qquad \tilde{K}^i_{\ j}=\frac{N}{\tilde{N}} \left( K^i_{\ j} +\frac{\dot{C}}{2N C}   \delta^i_{j} \right) \, .
\ee
For the matter sector, the stress-energy tensor in the new frame is
\be
\tilde T_{(I)}^{ \mu \nu} \equiv  \frac{2}{\sqrt{-\tilde g}} \frac{\delta S_I}{\delta \tilde  g_{\mu \nu}}  \;,
\ee
so that 
\be
\label{se_rel}
\tilde T_{(I)}^{\mu \nu} =   \frac{\sqrt{- g}}{\sqrt{- \tilde g}} \frac{\delta g_{\alpha \beta} }{\delta \tilde  g_{\mu \nu}} T_{(I)}^{\alpha \beta}  = \frac{N}{C^{5/2} \sqrt{CN^2-D}}\, T_{(I)}^{\mu \nu}  \;. 
\ee

\subsection{Background }
\label{app:bkg}

Let us now set $\bN=1$ and assume a flat FLRW background, $ds^2 = - dt^2 + a^2(t) d\mathbf{x}^2$.  From eq.~\eqref{disformalADM}, the background metric in the new frame remains flat FLRW, with line element $d\tilde s^2 =  - d\tilde t^2 + \tilde a^2 (\tilde t) d\mathbf{x}^2$, where we have defined 
\be
\label{time_a}
\tilde t \equiv \int \sqrt{\frac{C}{ 1+\alphaD}} dt \;, \qquad \tilde a \equiv \sqrt{C} a \;.
\ee
From this equation, the Hubble rate in the new frame is given by
\be
\label{Hchange}
\tilde H \equiv \frac{1}{\tilde a } \frac{d \tilde a}{d \tilde t} =  (1+\alphaC) \sqrt{  \frac{1+\alphaD}{C}} H \;.
\ee
From eq.~\eqref{se_rel}, the background energy density and pressure in the two frames are respectively related by
\be
\label{changeback}
\tilde \rho_{I} = \frac{1}{C^2\sqrt{1+ \alphaD}} \rho_I \;, \qquad  \tilde p_{I} = \frac{ \sqrt{1+ \alphaD}}{C^2} p_I \;,
\ee
where $\tilde \rho \equiv -\tilde T^0_{\ 0}$ and $\tilde p \equiv \tilde T^i_{\ i} /3 $. This implies $ \tilde w_I=w_I(1+\alphaD)$.

In terms of these quantities the Friedmann equations \eqref{Fr1} and \eqref{Fr2} become
\begin{align}
\tilde H^2 &=\frac{1}{3 \tilde M^2} (\tilde \rho_{\rm m}+\tilde \rho_{\rm DE}) \, ,\\
\frac{ d \tilde H}{d \tilde t}+
\frac32 \tilde H^2 
&=-\frac{1 }{2 \tilde M^2} (\tilde p_{\rm m} + \tilde p_{\rm DE} ) \, ,
\end{align}
where $\tilde M^2$ is given by eq.~\eqref{Mtilde}.
Using the expressions above, one can compute the relations between the background energy density and pressure of dark energy  in the two frames. One finds, respectively,
\begin{align}
\tilde \rho_{\rm DE}&=\frac{1}{C^2\sqrt{1+\alphaD}  } \rhoD +3M^2H^2 \frac{ \sqrt{1+\alphaD} }{C^2} \bigg[ \alphaC(2+\alphaC)+\frac{\alphaD}{1+\alphaD} \bigg]\, ,\label{defrhod}\\
\tilde p_{\rm DE} &= \frac{  \sqrt{1+\alphaD}}{C^2 } \pD+ M^2 H^2 \frac{\sqrt{1+\alphaD} }{C^2  } \bigg[\alphaC(4+\alphaC)+2\alphaC \frac{\dot H}{H^2} +2  \frac{\dot \alpha_{\rm C}}{H} + \frac{\dot \alpha_{\rm D} (1+\alphaC)}{H( 1+\alphaD)}\bigg]\, ,\label{defpd}
\end{align}
where {as usual} a dot denotes a derivative with respect to $t$.

\subsection{Perturbations}
\label{sec:B2}

Let us now study how perturbations transform under disformal transformations.
Due to the invariance of the gravitational action under disformal transformations, the perturbation equations  have the same form in both frames. 
Thus, we just need to derive the relation between perturbation quantities in different frames.

Introducing $\pi$ in eq.~\eqref{disf_uni_App} via the time reparametrization \eqref{stuek}, one finds, up to linear order in $\pi$,
\begin{align}
\tilde g_{00} &= C \bigg[  g_{00} + \frac{\alphaD}{1+\alphaD} (1+ 2 \dot \pi)  -  
 \frac{ 2  \alphaC }{1+\alphaD} H \pi + \frac{\dot \alpha_{\rm D}}{(1+\alphaD)^2} \pi   \bigg] \;,  \\\tilde g_{0i} & = C \bigg[ g_{0i} + \frac{\alphaD}{1+\alphaD} \partial_i \pi\bigg] \;, \\
\tilde g_{ij} & = C(1 + 2 H \alphaC \pi) g_{ij} \;.
\end{align}
Thus, if we start from a perturbed FLRW metric in Newtonian gauge with $g_{0i} = 0$ we end up with $\tilde g_{0i} \neq 0$ after this transformation. To maintain the Newtonian gauge condition $\tilde g_{0i} =0$, we need to supplement the time redefinition \eqref{time_a} with a space-dependent shift (see Appendix C of \cite{Gleyzes:2014qga}), i.e.
\be \label{tJ}
\tilde t =\int \sqrt{\frac{C}{1+\alphaD}} \, dt - \alphaD  \sqrt{\frac{C}{1+\alphaD}}  \,\pi(t,\mathbf{x}) \, .
\ee
Then, the {new perturbations}  result from combination of the field redefinition \eqref{disf_uni_App} and this change of coordinates. For the metric in Newtonian gauge, this yields
\begin{align}
\tilde  \Phi &=(1+\alphaD)\Phi+ \left[ \alphaC(1+\alphaD) H +\dot{\alpha}_{\rm D} \right]  \,\pi \label{eqPhiJ} \;, \\
\tilde  \Psi&= \Psi -[ \alphaC( 1+\alphaD)+ \alphaD] H \,\pi \label{eqPsiJ} \;, \\
\tilde  \pi &=\sqrt{C(1+\alphaD)}\,  \pi \label{eqPiJ}\, .
\end{align}
For the matter quantities, using eq.~\eqref{se_rel},  one finds
\begin{align}
\tilde \delta _{I}=&\ \delta_{I}+\alphaD\left(\Phi - \dot \pi\right) -\left[ 3 H \alphaD(1+w_{I} -\gamma_{I})+4 H \alphaC(1+\alphaD) + \frac12 {\dot{\alpha}_{\rm D}} \right]\pi\, , \label{rhotilderho}\\
 {\delta \tilde p_{I} }/{ \tilde \rho_I }=& \ (1+\alphaD)\bigg\{ {\delta p_{I}}/{\rho_I}-  w_{I} \alphaD \left(\Phi-\dot \pi\right) \nonumber \\
&- w_{I}  \left[3H \alphaD(1+w_{I}-\gamma_{I})+4H \alphaC(1+\alphaD) - \frac12 { \dot \alpha_{\rm D}} -   \frac{\dot w_{I}}{w_{I} }\alphaD \right]\pi\bigg\}\,,\label{ptildep}\\
\tilde v_{I}=&\ \frac{\sqrt{C(1+\alphaD)}}{1+w_{I}(1+\alphaD)}\left[(1+w_{I})v_{I}-w_{I}\alphaD\pi\right]\, ,\label{vtildev} \\
\tilde \sigma _{I}=&\ \frac{\sqrt{1+\alpha _D}}{ C } \,\sigma_{I}\, .
\end{align}
One can relate the pressure  and density perturbations via the speed of sound, which is defined as the ratio between these two quantities  in a coordinate system where the fluid is at rest. In the Jordan frame of the fluid, this {gives} \cite{Bean:2003fb}
\be
{ \delta \tilde p_I}= \tilde c_{s,I}^2 \delta \tilde \rho_I - (1+\tilde w_I) \tilde \rho_I \left[ 3 \tilde H(\tilde c_{s,I}^2-\tilde w_I) + \frac{d\tilde{w_I}/d \tilde t}{1+\tilde w_I}\right] \tilde v_I \;.
\ee
One can then use eqs.~\eqref{rhotilderho}--\eqref{vtildev} to rewrite this equation in  a generic frame. This yields
\be
\begin{split}
{\delta p_I}= &\ c_{s,I}^2  \delta \rho_I - \rho_I \left[3H(\tilde c_{s,I}^2-\tilde w_I)(1+\alphaC)+\frac{\dot{ \tilde{w}}_I}{1+\tilde w_I}\right]\left[(1+w_I)v_I+\frac{\alphaD}{1+\alphaD}\pi\right]\\\
&-4H \rho_I(c_{s,I}^2-w_I)\alphaC\pi+\rho_I (c_{s,I}^2+w_I)\left[2 \alphaD(\Phi-\dot \pi)-\frac{\dot \alpha_{\rm D}}{1+\alphaD}\pi\right]\, ,
\end{split}
\ee
where we {recall} that, from eqs.~\eqref{changeback} and \eqref{changecs}, the equation of state parameters and sound speeds defined in the two frames are respectively related by
$\tilde w_I = (1+\alphaD){ w}_I$ and  $\tilde c_{s,I}^2= (1+\alphaD) { c_{s,I}^2} $.

\section{Explicit quadratic action}
\label{Lagmatter}
\subsection{Matter action}
For simplicity, we assume that each matter species can be described by a perfect fluid with vanishing vorticity (this restriction does not affect the analysis of {\it scalar} linear modes). It is then easy to write an action in terms of a derivatively coupled scalar field with Lagrangian\footnote{The more general $k$-essence type Lagrangian of Refs.~\cite{ArmendarizPicon:2000dh,ArmendarizPicon:2000ah} explicitly depends also on the scalar field. Since here we are interested only in the derivative terms, we assume for simplicity that   $P_I$ depends only on $Y_I$ and not on $\sigma_I$.  This description implies that each of the fluids is also barotropic \cite{Boubekeur:2008kn}, i.e.~that its pressure is a function of its energy density, $p_I = p_I(\rho_I)$.}
\be
\label{Lagmatkess}
S_{\rm m} = \sum_I^N S_I \;, \qquad S_I= \int d^4 x \sqrt{- \tg^{(I)}} P_I(Y_I ) \;, \qquad Y_I \equiv  \tg_{(I)}^{\mu \nu} \partial_\mu \sigma_I \partial_\nu \sigma_I \;.
\ee
The second-order expansion of the action $S_I$ reads
 \be
 \label{actkessence}
 \begin{split}
 S^{(2)}_I  = & \int d^3 x \, dt \,\bN \frac{ a^3}{\cI^2} \left\{ \frac{1   +  \alphaDI \,  \cI^2 + (1+\alphaDI) w_I }{ 2 } \rho_I \left( \frac{\delta N}{\bN} \right)^2  \right. \\
 & - \frac{1+(1+\alphaDI) w_I}{(1+\alphaDI)^2} \frac{ \rho_I }{  \dot \bs_I} \left[ \delta \dot \sigma_I \left(\frac{\delta N}{\bN} - \cI^2 \delta \sqrt{h}\right) + \cI^2 N^i \partial_i \delta \sigma_I \right] \\
 & + \left. \frac{1+(1+\alphaDI) w_I}{(1+\alphaDI)^2} \frac{ \rho_I }{ 2  \dot \bs_I^2} \left[ \delta \dot \sigma_I^2 - \bN^2 \cI^2 \frac{ (\partial_i \delta \sigma)^2}{a^2} \right] \right\} \;,
 \end{split}
\ee
where we have split  the scalar field {$\sigma_I$} into a background value and its perturbations, $\sigma_I = \bs_I (t) + \delta \sigma_I (t, \mathbf{ x}) $. The fluid quantities are related to the function $P_I\left(Y_I\right)$ through
\be
\begin{split}
p_I &\equiv \frac{C_I^2}{\sqrt{1+\alphaDI}}P_I \, , \qquad \rho_I\equiv C_I^2 \sqrt{1+\alphaDI}\left(2Y_I P'_{I}-P_I\right)\, , \\
\cI^2& \equiv \frac{P'_{I}}{P'_{I} +2 Y_I  P''_{I} }\left(1+\alphaDI\right)^{-1} \;,
\end{split}
\ee
where a prime denotes a derivative with respect to the variable $Y_I$. We have omitted in the action irrelevant terms that vanish when imposing the background equations of motion. 
For $C_I =1$ and $\alphaDI=0$ we recover the usual expressions for a $k$-essence fluid \cite{ArmendarizPicon:2000dh,ArmendarizPicon:2000ah,Garriga:1999vw}.

\subsection{Stability and sound speed of dark energy}

In order to investigate linear stability issues, we need to extract the quadratic action for the propagating degrees of freedom. We concentrate on scalar modes as the stability conditions of tensors are not modified by the nonminimal coupling of matter. To this end, we will expand the total  action up to quadratic order in  
linear scalar fluctuations around a FLRW solution and solve the constraints,  generalizing the procedure of Refs.~\cite{Gergely:2014rna} and \cite{Gleyzes:2014qga}. 

The second-order action
\be
\label{sec_tot_ac}
S^{(2)} = S^{(2)}_{\rm g} + S^{(2)}_{\rm m} \;, 
\ee
where the gravitational part $S^{(2)}_{\rm g}$ is given in eq.~\eqref{S2},
governs the dynamics of linear scalar fluctuations.
Assuming $\bN=1$ without loss of generality, 
the scalar modes can be described in unitary gauge
by the 
metric
perturbations  
\cite{Maldacena:2002vr}
\be
N=1+\d N, \quad N^i=\delta^{ij}\partial_j \psi, \quad {h}_{ij}=a^2(t) e^{2\zeta}\delta_{ij}\,. \label{metric_ADM_pert}
\ee
As a consequence, we get 
\be
\label{dhdK}
\d\sqrt{h} = 3 a^3 \zeta\,, \qquad  \d K^i_{\ j}=\big(\dot\zeta-H\d N\big)\d^i_j-\d^{ik}\partial_{k}\partial_j\psi \;,
\ee
and
\be \label{RRR}
\d_1\sR_{ij}
=  - \delta_{ij} \partial^2 \zeta -  \partial_i \partial_j \zeta \;, \qquad \d_2 \sR=  -\frac{2}{a^2}\left[(\partial\zeta)^2-4\zeta\partial^2\zeta\right]\,.
\ee
(The metric perturbations $\delta N$ and $\zeta$ and the scalar fluctuation $\psi$ are related to the metric perturbations in Newtonian gauge by
$\delta N = \Phi -\dot \pi$, $\zeta= -\Psi   -  H \pi $ and  $\psi = a^{-2} \pi$.)
Substituting these expressions
into  (\ref{sec_tot_ac}), we obtain the second-order action in terms of the three scalar quantities $\delta N$, $\psi$ and $\zeta$. 
Variation with respect to $\psi$ yields the momentum constraint, whose  solution reads
\be 
\delta N =  \frac{1}{1+\alphaB}\left(\frac{\dot \zeta}{H} + \frac32H\sum_I \frac{1+(1+\alphaDI) w_I}{1+\alphaDI }\Omega_I \frac{\delta\sigma_I}{\dot\sigma_I} \right) \;,
\ee
with $\Omega_I =  {\rho_I}/ ({3M^2 H^2})$.

We do not need the solution of the Hamiltonian constraint, as the longitudinal part of the shift $\psi$ only contributes to a boundary term in the action. Replacing the above solution into the second-order action and re-expressing the scalar fields perturbations $\delta \sigma_I$ in terms of the gauge invariant variables
\be
\Q_I \equiv \delta \sigma - \frac{\dot \bs_I}{H} \zeta \;,
\ee
the total second-order action reads, focusing only on the {kinetic and spatial gradient parts},
\be
\label{QuadActI}
\begin{split}
S^{(2)} = &\int d^3 x \, dt \, a^3  \, \frac{M^2}{2  }   \bigg[  g_{\dot \zeta \dot \zeta}  \dot \zeta^2 + g_{\partial \zeta \partial \zeta} \frac{(\partial_i \zeta)^2}{a^2}   
+ \sum_I  \frac{\kappa_I H^2 }{\dot \bs_I^2\cI^2} \left(  \dot \Q_I^2    - \cI^2 \frac{(\partial_i \Q_I)^2}{a^2} \right) \\
& +2 \sum_I  \frac{g_{{\rm int},I} H}{\dot \bs_I}  \left( {\dot \Q_I} \dot \zeta   - \frac{\cI^2}{a^2}  {\partial_i \Q_I} \partial_i \zeta \right)  
\bigg] \;,
\end{split}
\ee
with
\begin{align}
g_{\dot \zeta \dot \zeta} &\equiv \frac{1}{(1+\alphaB)^2}\left[\DDt+  \sum_I  \frac{\kappa_I}{\cI^2} (\alphaDI-\alphaB)^2
\right] \;, \\
g_{{\rm int},I} &\equiv  \frac{1}{1+\alphaB}\sum_I  \frac{\kappa_I}{\cI^2} (\alphaDI-\alphaB) \;, \\
g_{\partial \zeta \partial \zeta} &\equiv 
\frac{2}{1+\alphaB}\left[\frac{\dot H}{H^2}+\frac{\dot\alpha_{\rm B}}{1+\alphaB}+\alphaB (1+\alphaT)+\alphaT-\alphaM+ \sum_I  \frac{\kappa_I}{2} (1+2\alphaDI-\alphaB)\right]
\;,
\end{align}
where we have defined the dimensionless coefficients
\be
\DDt \equiv \alphaK + 6 \alphaB^2 +3 \sum_I \alphaDI \, \Omega_I \;, \qquad 
\kappa_I \equiv 3 \frac{1+(1+\alphaDI) w_I}{(1+\alphaDI)^2} \Omega_I \;. \label{kappa_def}
\ee
Absence of ghosts is ensured by requiring that the matrix of the kinetic coefficients is positive definite, which yields the conditions $\DDt \geq 0$ and $\kappa_I \geq 0$. The second condition reads $\rho_I+(1+\alphaDI) p_I\geq0$, which is  the usual Null Energy Condition written in a disformed frame.

Diagonalization of the {kinetic-spatial gradient} matrix yields the following speed of propagation for dark energy,
\be
\label{ssApp}
c_s^2 = -\frac{2}{\DDt}\bigg\{ (1+\alphaB) \bigg[ \frac{\dot H}{H^2} -\alphaM + \alphaT  +  \alphaB (1+\alphaT)\bigg] + \frac{\dot \alpha_{\rm B}}{H} +\frac32 \sum_I  \Big[ 1+ (1+\alphaDI) w_I \Big] \Omega_I\bigg\} \;.
\ee
Absence of gradient instabilities requires $c_s^2 \ge 0$ and $\cI^2 \ge 0$. 

\section{Perturbation equations}
\label{app1}

Here we provide the generalized Einstein equations in the presence of dark energy and modifications of gravity. These have been first given in Ref.~\cite{Gleyzes:2013ooa} in terms of the parameters of the Effective Field Theory of dark energy \cite{Gubitosi:2012hu} and in Refs.~\cite{Bellini:2014fua} (see also \cite{Gleyzes:2014rba}) in terms of the parameters $\alpha_a$.

\subsection{Einstein equations}

Let us defined 
\be
w_{\rm m} \equiv \sum_I \frac{\rho_I}{\rho_{\rm m}} w_I \;, \qquad \gamma_{\rm m} \equiv \sum_I \frac{\rho_I}{\rho_{\rm m}} \gamma_I \;,
\ee
where $\gamma_I$ parametrizes the nonminimal coupling of the species $I$, see definition in eq.~\eqref{gammab}.
The Hamiltonian constraint ((00) component of the Einstein equation) is
\be
\begin{split}
&6 (1+\alphaB)H \dot \Psi + (6-\alphaK +12 \alphaB )H^2  \Phi + 2  \frac{k^2}{a^2}\Psi +  \left(\alphaK -6 \alphaB \right)H^2 \dot \pi  \\
& +6 \left[  (1+\alphaB) \dot H + \frac32 H^2 \Omega_{\rm m} (1 + w_{\rm m} - \gamma_{\rm m})  - \frac13 \frac{k^2}{a^2} \alphaB \right]  H  \pi = -3\Omega_{\rm m}H^2 \delta_{\rm m} \;,  \label{Ein1}
\end{split}
\ee
while the momentum constraint ((0$i$) components of the Einstein equation) reads
\be
2 \dot \Psi + 2(1+\alphaB) H \Phi -2 H \alphaB \dot \pi +  \big[ 2\dot H  +3H^2\Omega_{\rm m}(1 + w_{\rm m}) \big] \pi= - 3H^2\Omega_{\rm m}(1 + w_{\rm m}) v_{\rm m} \;.\label{Ein2}
\ee
The traceless part of the $ij$ components of the Einstein equation gives
\be
 \Phi- (1+\alphaT) \Psi + (\alphaM - \alphaT) H\pi = -\frac{\sigma_{\rm m}}{M^2}  \;,\label{Ein3}
\ee
while the trace of the same components gives,
using the  equation above,
\be
\begin{split}
&2 \ddot\Psi + 2 (3 + \alphaM) H \dot  \Psi+ 2 (1 + \alphaB ) H \dot \Phi   \\
& + 2 \left[   \dot H  - \frac{3}{2}{H^2\Omega_{\rm m}(1 + w_{\rm m})}  +  ( \alphaB H)^{\hbox{$\cdot$}}+  (3 + \alphaM)(1 + \alphaB) H^2  \right] \Phi \\
&-2H\alphaB\, \ddot \pi+ 2 \left[    \dot H + \frac{3}{2}{H^2\Omega_{\rm m}(1 + w_{\rm m})}  -  ( \alphaB H)^{\hbox{$\cdot$}} - (3 + \alphaM) \alphaB H^2  \right] \dot \pi \\
& +2 \left\{  (3+\alpha _M )H \dot H+ \frac32 H^2 \Omega_{\rm m} {\left[ \dot w_{\rm m}-3H(1+w_{\rm m} -\gamma_{\rm m} )\right]} + \ddot H \right\} \pi = \frac{1}{M^2 } \left(\delta p_{\rm m} - \frac{2}{3} \frac{k^2}{a^2} \sigma_{\rm m} \right)\;.\label{Ein4}
\end{split}
\ee

\subsection{Scalar field equation} 

The charge $Q_I$ is defined in eq.~\eqref{def_QI}.
Its perturbation reads
\be
\label{Qpert}
\begin{split}
\delta Q_I & \equiv 3 H \left[ \gamma_I \delta_I - \alphaCI \left(\frac{\delta p_I}{\rho_I } - w_I \delta_I \right)\right] \rho_I - \alphaDI \frac{1+w_I}{1+ \alphaDI} \frac{k^2}{a^2} \rho_I {v_I} \\
& + 2H \big[ \alphaCI (1-3 w_I) - 3 \gamma_I  \big] \rho_I \Phi -\frac{\alphaDI}{1+ \alphaDI} \left( \dot \Phi + 3 \dot \Psi - \dot \delta_I - \ddot \pi + w_I \frac{k^2}{a^2} {\pi} \right) \rho_I \\
& +\frac{H}{1+\alphaDI } \big[ - 2 \alphaCI (1 - 3 w_I)(1+\alphaDI ) + 3 w_I \alphaDI + 3 \gamma_I (2+ \alphaDI )\big] \rho_I \dot \pi \\
&+ \frac{3}{1+\alphaDI} \Big\{ \big( w_I \alphaDI + \gamma_I \big) \dot H + \big[(\alphaDI + \alphaCI(1+\alphaDI)) \dot w_I+\dot \gamma_I   \big] H \Big\} \rho_I {\pi} \;.
\end{split}
 \ee
We have checked that this expression agrees with those in the literature (see e.g.~\cite{vandeBruck:2012vq,Koivisto:2012za,Zumalacarregui:2012us,vandeBruck:2015ida,Minamitsuji:2014waa}) in the relevant limits.\footnote{The expressions for the charge $Q_I$ given in eqs.~\eqref{barQ} and \eqref{Qpert} are in unitary gauge. To compare  to those in the literature, one must rescale by a factor $\dot{\phi}$, i.e.~$Q_I\mapsto Q_I\dot{\phi}$ and $\delta Q_I\mapsto \delta Q_I\dot{\phi}+ \bar{Q}_I ({\ddot{\phi}}/{\dot{\phi}}) \pi$.}

The evolution equation for $\pi$ in the absence of EP violations is given in \cite{Gleyzes:2013ooa} and can be found in \cite{Gleyzes:2014rba} in terms of the parameters used in this article. Including the contribution of $ \sum_I \delta Q_I$ using the above equation, and using the continuity equation, eq.~\eqref{continuity_pert}, this becomes
\be
\begin{split}
&\Big(\alphaK +3 \sum_I\alphaDI  \Omega_I \Big) H^2 \ddot \pi +\bigg\{ \left[ H^2(3+\alphaM)+\dot H \right] \alphaK+(H\alphaK)^{\hbox{$\cdot$}} \\
& -3H^2 \sum_I \Omega_I
\Big[2 \alphaCI (1 - 3 w_I)(1+\alphaDI) -3 w_I \alphaDI - 6 \gamma_I (1+\alphaDI) \Big] \bigg\} H\dot \pi \\
&+3\bigg\{ 2 \dot H^2+ 3 \dot H H^2 \bigg[ \Omega_{\rm m} (1+w_{\rm m}) + \sum_I  w_I \alphaDI  \Omega_I \bigg]  +2\dot H \alphaB \left[ H^2(3+\alphaM)+\dot H \right]
\\
&+2H(\dot H \alphaB)^{\hbox{$\cdot$}}   { + 3H^3\sum_I \Big[  \dot w_I \left(\alphaDI + \alphaCI(1+\alphaDI)\right)  + 3 H \gamma_I(1+ w_I - \gamma_I) \Big] \Omega_I }
\bigg\} \pi \\
&-\frac{k^2}{a^2}\bigg\{ 2 \dot H+ 3 H^2 \Omega_{\rm m} (1+w_{\rm m})    + 2 H^2 \Big[ \alphaB(1+\alphaM)+\alphaT -\alphaM \Big]
+2 \left(H\alphaB\right)^{\hbox{$\cdot$}} \\&
+3H^2 \sum_I w_I\alphaDI \Omega_I \bigg\} \pi  + 6H\alphaB \ddot \Psi + \Big[ H^2(6\alphaB-\alphaK) -3H^2 \sum_I \alphaDI \Omega_I\Big] \dot \Phi \\
&+3\Big[2  \dot H+ 3 H^2 \Omega_{\rm m} (1+w_{\rm m})  +2 H^2\alphaB(3+\alphaM) + 2 (\alphaB H)^{\hbox{$\cdot$}} +3H^2 \sum_I w_I \alphaDI \Omega_I \Big] \dot \Psi \\
&+\bigg\{6 \dot H+ 9  H^2 \Omega_{\rm m} (1+w_{\rm m})  +H^2(6\alphaB-\alphaK)(3+\alphaM)+2(9\alphaB-\alphaK)\dot H+H(6\dot\alphaB-\dot \alphaK) \\
&- 6H^2 \sum_I \Big[ 3 \gamma_I - \alphaCI (1-3 w_I) \Big](1+ \alphaDI) \Omega_I \bigg\}H\Phi 
+2\frac{k^2}{a^2} \left\{ \left[H(\alphaM-\alphaT) \right]\Psi -\alphaB H\Phi \right\} \\
& +9H^3 \sum_I  \left\{  \gamma_I \delta_I - \left[\alphaCI(1+\alphaDI) + \alphaDI \right]\left(\frac{\delta p_I}{\rho_I}- w_I\delta_I \right) \right\} \Omega_I =0 \;. \label{sfee}
\end{split}
 \ee

\section{Synchronous gauge}
\label{app_sync}

Here we provide the perturbation equations in synchronous gauge, {often employed in numerical codes,}
where the perturbed FLRW metric has the form
\be
ds^2= -d t^2+ \bigg[ \bigg( 1 + \frac{1}3h \bigg) \delta_{ij}+\left( \frac{ k_i k_j}{k^2} -\frac{1}3\delta_{ij} \right) \left(h+6\eta\right) \bigg] dx^idx^j \, .
\ee
Defining $\epsilon \equiv {a^2 \big(  \dot h +6  \dot \eta \big) / k^2}$, one can write Newtonian gauge quantities in terms of synchronous gauge ones using the following relations (see for instance \cite{Ma:1995ey}),
\be
\begin{split}
\Phi &= \dot  \epsilon \, , \qquad
\Psi =\eta-H \epsilon\, ,\qquad \pi^{(N)} =\pi^{(S)}+ \epsilon \;, \\
 \delta \rho_I^{(N)}&= \delta \rho_I^{(S)}  + \dot \rho_I  \epsilon \, ,\qquad \delta p_I^{(N)}  = \delta p_I^{(S)}+ \dot p_I  \epsilon \, , \quad
\quad v_I^{(N)} =- {\theta_I^{(S)}}/{k^2}-\epsilon \,, 
\end{split}
\ee
where we have introduced the divergence of the velocity, {$\theta_I \equiv - k^2  v_I/a$}. (The anisotropic stress is  gauge invariant.) We can then use the above relations to rewrite eqs.~\eqref{Ein1}--\eqref{Ein4} in synchronous gauge.
To do this, we use conformal time, $\tau \equiv \int dt/a$, and  denote by a prime the derivative with respect to it.
Rescaling the scalar fluctuation $\pi$ by the conformal factor, $\pi \to {\pi}/a$, 
and defining the conformal Hubble rate as ${\cal H} \equiv {a'}/{a}$,  one obtains\\
($(00)$ component)
\be
\begin{split}
&2k^2 \eta-{\cal H}(1+\alphaB)h'-{\cal H}^2(6\alphaB-\alphaK)\pi'+\Big[ 9 {\cal H}^2 \Omega_{\rm m} (1+w_{\rm m} 
- \gamma_{\rm m})
\\
&-{\cal H}^2( 6 - \alphaK+12\alphaB)+6{\cal H}' (1+\alphaB)  - 2k^2 \alphaB \Big]{\cal H} \pi=-\frac{a^2}{M^2} \rho_{\rm m} \delta_{\rm m} \, ,
\end{split}
\ee
($(0i)$ component)
\be
\begin{split}
2 \eta'- 2 {\cal H}\alphaB\pi'   + \Big[ 2 {\cal H}' - 2(1+\alphaB)  {\cal H}^2 + 3 {\cal H}^2 \Omega_{\rm m} (1+w_{\rm m}) \Big] \pi =\frac{a^2}{M^2}(\rho_m+p_m) \frac{\theta_{\rm m}}{k^2}\, ,
\end{split}
\ee
($(ij)$-traceless)
\be
\begin{split}
h''+6\eta''+{\cal H}(2+\alphaM)(h'+6\eta')-2k^2(1+\alphaT)\eta  -2k^2{\cal H}\left(\alphaT-\alphaM\right) \pi= \frac{2k^2}{M^2}\sigma_{\rm m}\, ,
\end{split}
\ee
and 
($(ij)$-trace)
\be
\begin{split}
&h''+{\cal H}(2+\alphaM)h'-2k^2(1+\alphaT)\eta +6\alphaB {\cal H}\pi'' \\
&+\Big[6{\cal H}^2\alphaB(3+\alphaM)+6 (\alphaB{\cal H})'  - 9 {\cal H}^2 \Omega_{\rm m} (1+w_{\rm m}) - 6 ({\cal H}' - {\cal H}^2) \Big]\pi' \\
&+\Big\{6{\cal H}^2[ 2 +  \alphaM+ \alphaB(2+\alphaM) ] +6 (\alphaB  -\alphaM) {\cal H}'+6(\alphaB {\cal H})'-2k^2( \alphaT-\alphaM)\\
& - 9 {\cal H}^2 \Omega_{\rm m}  \Big[ (1- 3 w_{\rm m}) (1+w_{\rm m})+ 3 w_{\rm m} \gamma_{\rm m} + w_{\rm m}'/{\cal H} \Big]  - 6 {\cal H}''  \Big\}{\cal H}\pi =- 3 \frac{a^2}{M^2} \delta p_{\rm m}  \, .
\end{split}
\ee

In synchronous gauge, the evolution equation for the scalar fluctuation, eq.~\eqref{sfee}, reads
\be
\begin{split}
&{\cal H}^2\Big(\alphaK +3\sum_I\alphaDI  \Omega_I \Big)\pi''+\bigg\{{\cal H}^2\alphaK(2+\alphaM)+{\cal {\cal H}}'\alphaK+(\alphaK{\cal H})' +\\
&-3{\cal H}^2\sum_I \Omega_I  \Big[ 2\alphaCI (1-3w_I)(1+\alphaDI)-\alphaDI (1+3w_I)-6 \gamma_I  (1+\alphaDI) \Big] \bigg\}{\cal H}\pi'\\
&-2k^2\bigg\{{\cal H}^2\big[ \alphaB \alphaM+\alphaT-\alphaM - 1\big]+\left( \alphaB{\cal H}\right)'   + {\cal H}'  +\frac{3}{2}\sum_I{\cal H}^2 \Big[ 1 + w_I (1+\alphaDI) \Big] \Omega_I \bigg\}\pi
\\& + \bigg\{{\cal H}^4[ 6 -6\alphaB\alphaM + \alphaK(1+\alphaM) ]
+3{\cal H}^2{\cal H}' [ -4 + \alphaK-2\alphaB(3-\alphaM)]\\
&+6(1+\alphaB) {\cal H}'^2-{\cal H}^3(6\alphaB'-\alphaK')+6{\cal H} ( \alphaB{\cal H}' )'    \\
&+3{\cal H}^2 \sum_I \Omega_I \bigg[-2{\cal H}^2 \alphaCI (1-3w_I)(1+\alphaDI)    + {\cal H}' \Big(3 + \alphaDI +3w_I (1+\alphaDI)  \Big)    \\
&+3 {\cal H}^2 \Big( \gamma_I (1+2\alphaDI) -  (1 -3 \gamma_I) (1+ w_I  - \gamma_I )\Big)  +3 {\cal H}  w_I'  \Big(  \alphaDI+\alphaCI(1+\alphaDI)  \Big)\bigg]\bigg\}\pi\\
&-{\cal H}\alphaB h''   -\bigg\{ {\cal H}'  + {\cal H}^2 \Big[ \alphaB(1+\alphaM) - 1 \Big] +(\alphaB {\cal H})' +\frac{3}{2}\sum_I{\cal H}^2 \Big[ 1 + w_I (1+\alphaDI)  \Big] \Omega_I   \bigg\} h' \\
&+2k^2  {\cal H} (\alphaM-\alphaT)  \eta   =0 \, .
\end{split}
\ee
The continuity and Euler equations for matter become, {respectively,}
\be
\begin{split}
&\delta_I'+3{\cal H}  (1+\alphaCI) (1+\alphaDI)\bigg(\frac{\delta p_I}{\rho_I} - w_I\delta_I\bigg) +(1+w_I) \theta_I-\alphaDI  \pi''   \\
&- {\cal H} \big[-2 \alphaCI(1-3 w_I) (1+\alphaDI) +\alphaDI (1+3 w_I+6\gamma_I)+9 \gamma_I\big] \pi' \\
& +\bigg\{ -2 {\cal H}^2 (1+\alphaDI) \big( 3 \gamma_I-\alphaCI (1-3w_I)\big)+w_I \alphaDI k^2 \\
&-{\cal H}' \big(3\gamma_I+\alphaDI(1+3w_I) \big)-3 {\cal H} \Big[ w_I'\big(\alphaDI+\alphaCI(1+\alphaDI)\big)+\gamma_I'\Big]\bigg\} \pi \\
&+\frac{1}{2}\Big[1+w_I (1+\alphaDI)\Big] h'=0 \, ,
\end{split}
\ee
and
\be
\begin{split}
&\theta_I'+{\cal H} \bigg[ 1-3w_I+3\gamma_I+\frac{w_I'}{ {\cal H}(1+w_I)}\bigg] \theta_I-\frac{k^2 \delta p_I}{\rho_I (1+w_I)}+\frac{2 k^4}{3 a^2 \rho_I (1+w_I)} \sigma_I =\frac{3 {\cal H} \gamma_I}{(1+w_I)} k^2 \pi \, .
\end{split}
\ee

\section{Definitions of the parameters}
\label{parameters}

The coefficients $\nbeta_a$ appearing in eqs.~\eqref{Phi_evol} and \eqref{Phitopsi} are defined as
\begin{align}
\mbeta_1  & \equiv  - \frac{3  }{4 }  \Omega_{\rm m} \alphaK - \frac12 \DD \bigg( \frac{\dot H}{H^2} + \alphaT -  \alphaM   \bigg)-\frac{9}2\alphaB\gammac \Omega_c  \;, \label{beta1}\\
 \frac{\mbeta_1 \nbeta_2}{\DD}& \equiv\frac{9}{2} \Omega_{\rm m} \alphaB \left[ \frac{ \alphaB}{\DD} \nbeta_3 - \frac{4+\alphaM+\alphaT}{6\alphaB} +\frac{\xi}{\DDt} \right]-\frac94\gammac \Omega_c\left( c_s^2-\frac{2\alphaB\beta_3}{\DD}-2\alphaB\frac{3-3\gammac- \xi}{\DDt}\right) \nonumber 
\\ & + \frac12 (1+\alphaM) \bigg[ (\alphaM-\alphaT)-\frac{\dot H}{H^2} \bigg] -\frac{1}{2} \bigg[\frac{\dot \alphaM-\dot \alpha_{\rm T}}{H}+\frac{2\dot H(\alphaM-\alphaT)}{H^2}-\frac{\ddot H}{H^3}\bigg] \, ,\\
\nbeta_3 &\equiv3+\alphaM\frac{\DD}{\DDt}-6\frac{(\alpha_{\text{C},c}-3\gammac) (1+\alpha_{\text{D},c})+\alpha_{\text{D},c}}{\DDt} \Omega_c +\frac{\alphaB^2}{H  \DDt} \left(  \frac{ \alphaK}{\alphaB^2} \right)^{\hbox{$\cdot$}} \nonumber \\
&+3\Omega_c\frac{\alpha_{\text{D},c} }{\alphaB\DDt }\bigg[\frac{\dot H}{H^2}-\alphaM+(1+\alphaB)\alphaT-\frac{\left(\alphaB H\right)^{\hbox{$\cdot$}}}{H^2}+\frac{3}{2} \Omega_{\rm m} \bigg]\, ,  \\
\nbeta_4 & \equiv (1+\alphaT) \big( \nbeta_2-1-\alphaM+2\dot H/H^2 \big) + \dot \alphaT /H \;,  \\
\nbeta_5 & \equiv c_s^2 - \frac{ 2 \alphaB (\beta_3- \beta_2)}{\DD}
+\frac{\alphaB^2 }{\beta_1} (1+\alphaT) (\beta_3 - \beta_2 ) +\frac{\alphaB^2 \beta_4}{\beta_1}
 \;, \\
\nbeta_{6,I} & \equiv \nbeta_{7,I} + 2 \frac{\alphaB(\nbeta_2 -\nbeta_3)}{\DD} \;, \\
\nbeta_{7,I} & \equiv c_s^2+2\frac{ \alphaB\, \xi}{\DDt} \;, \\
\nbeta_{8,I} & \equiv \nbeta_{9,I} -(6\alphaB-\alphaK) \frac{\nbeta_2 -\nbeta_3}{\DD} \;, \\
\nbeta_{9,I} & \equiv -( 4+3c_s^2+\alphaM+\alphaT)+\beta_3 \;, 
\end{align}
where we remind that $\xi \equiv   \alphaB (1+\alphaT) + \alphaT - \alphaM$, $\DDt = \alphaK + 6 \alphaB^2 +3\alpha_{\text{D},c} \Omega_{c}$ and we have defined $\DD \equiv \alphaK + 6 \alphaB^2 = \DDt - 3\alpha_{\text{D},c} \Omega_{c}$. Setting $\alphaDc=\alphaCc=\gammac=0$ in these equations, one recovers the expressions of \cite{Bellini:2014fua} and \cite{Gleyzes:2014rba}.

For $\alphaM=\alphaT=0$, the coefficients $\xi_a$ appearing in eqs.~\eqref{conteqc2} and \eqref{Eulerc2} are defined as
\begin{align}
\xi_2 &\equiv \frac32 \frac{\DD}{\mbeta_1} \bigg[ \frac{\DDt}{\DD}\frac{ \left(H\gamma\right)^{\hbox{$\cdot$}}}{H^2} - \frac{\dot H}{H^2}\bigg] + \frac92 \frac{ \Omega_{\rm m}}{\mbeta_1}\left[\alphaB \Xi -\frac12 \DD -\alpha_{\text{D},c}\left(\frac{\DD}{\DDt}\alphaB+\frac{\alphaK}2c_s^2\right)\right]\nonumber \\
&-\frac{9\alphaDc \Omega_c }{2H^2\beta_1}\left(H\gamma\right)^{\hbox{$\cdot$}}+\frac{9\gammac \Omega_c}{2\beta_1}\left[3\alphaB\alphaDc\, c_s^2+\Xi+\frac{\alphaB+3\gammac-3}{\DDt}\DD\,\alphaDc\right]\, , \\
\xi_3&\equiv \alphaB \Xi -\frac{  \alphaK\, \alpha_{\text{D},c}}{2}\, c_s^2\, , \\
\xi_4&\equiv \xi_2-3(1-\gammac)\, , \\
\xi_5&\equiv \frac{\alphaB}{\mbeta_1}\Xi+2\frac{\alphaB\,\xi_2-\Xi}{\DD}-3\frac{\alphaB^2}{\beta_1}(1-\gammac)
-\left[\frac{\alphaK}{2\mbeta_1}+6\frac{\alphaB}{\DD}\right]\alpha_{\text{D},c} c_s^2\, , \\
\xi_{6,I}&\equiv -2\alphaB\, \alpha_{\text{D},c}\left(3\frac{c_s^2}{\DD}+\frac1{\DDt}\right)+2\frac{\alphaB\, \xi_2-\Xi}{\DD}  -6\frac{\alpha_{\text{D},c}}{\DDt} \gamma_I\, , \\
\xi_{7,I}&\equiv -\alphaB\, \alpha_{\text{D},c}\left(c_s^2+2\frac{\alphaB^2}{\DDt}\right)   -6\alphaB^2 \frac{\alpha_{\text{D},c}}{\DDt} \gamma_I \, , \\
\xi_{8,I}&\equiv-3+\frac{6\alphaB-\alphaK}{\DD}\left(3\alpha_{\text{D},c} c_s^2-\xi_2\right)+6\frac{1+\alphaB}{\DD}\Xi \, , \\
\xi_{9,I}&\equiv-3\alphaB^2+\frac{6\alphaB-\alphaK}2\alpha_{\text{D},c} c_s^2+\alphaB\Xi\, , 
\end{align}
with 
\be
\begin{split}
\Xi  \equiv & \ 3\alphaB+3\gammac - \frac{2\DD}{\DDt}\, \left[\alpha_{\text{D},c}+\alpha_{\text{C},c}(1+\alpha_{\text{D},c})-3\gammac(1+\alpha_{\text{D},c})+\alpha_{\text{D},c}\dot H/(2H^2)\right]  \\
&-\frac{\alpha_{\text{D},c}}{\DDt H^2}\left\{(1+\alphaB)\alphaK H^2+(\DD H)^{\hbox{$\cdot$}}-6\alphaB\left[(1+\alphaB)H\right]^{\hbox{$\cdot$}}-9\alphaB\Omega_{\rm m}H^2\right\}\, .  
\end{split}
\ee

\bibliographystyle{utphys}
\bibliography{EFT_DE_biblio}

\providecommand{\href}[2]{#2}\begingroup\raggedright\begin{thebibliography}{10}

\bibitem{Amendola:2012ys}
{\bf Euclid Theory Working Group} Collaboration, L.~Amendola {\em et.~al.},
  ``{Cosmology and fundamental physics with the Euclid satellite},'' {\em
  Living Rev.Rel.} {\bf 16} (2013) 6,
  \href{http://xxx.lanl.gov/abs/1206.1225}{{\tt 1206.1225}}.

\bibitem{Clifton:2011jh}
T.~Clifton, P.~G. Ferreira, A.~Padilla, and C.~Skordis, ``{Modified Gravity and
  Cosmology},'' {\em Phys.Rept.} {\bf 513} (2012) 1--189,
  \href{http://xxx.lanl.gov/abs/1106.2476}{{\tt 1106.2476}}.

\bibitem{Joyce:2014kja}
A.~Joyce, B.~Jain, J.~Khoury, and M.~Trodden, ``{Beyond the Cosmological
  Standard Model},'' {\em Phys.Rept.} {\bf 568} (2015) 1--98,
  \href{http://xxx.lanl.gov/abs/1407.0059}{{\tt 1407.0059}}.

\bibitem{Gubitosi:2012hu}
G.~Gubitosi, F.~Piazza, and F.~Vernizzi, ``{The Effective Field Theory of Dark
  Energy},'' {\em JCAP} {\bf 1302} (2013) 032,
  \href{http://xxx.lanl.gov/abs/1210.0201}{{\tt 1210.0201}}.

\bibitem{Gleyzes:2013ooa}
J.~Gleyzes, D.~Langlois, F.~Piazza, and F.~Vernizzi, ``{Essential Building
  Blocks of Dark Energy},'' {\em JCAP} {\bf 1308} (2013) 025,
  \href{http://xxx.lanl.gov/abs/1304.4840}{{\tt 1304.4840}}.

\bibitem{Creminelli:2006xe}
P.~Creminelli, M.~A. Luty, A.~Nicolis, and L.~Senatore, ``{Starting the
  Universe: Stable Violation of the Null Energy Condition and Non-standard
  Cosmologies},'' {\em JHEP} {\bf 0612} (2006) 080,
  \href{http://xxx.lanl.gov/abs/hep-th/0606090}{{\tt hep-th/0606090}}.

\bibitem{Cheung:2007st}
C.~Cheung, P.~Creminelli, A.~L. Fitzpatrick, J.~Kaplan, and L.~Senatore, ``{The
  Effective Field Theory of Inflation},'' {\em JHEP} {\bf 0803} (2008) 014,
  \href{http://xxx.lanl.gov/abs/0709.0293}{{\tt 0709.0293}}.

\bibitem{Creminelli:2008wc}
P.~Creminelli, G.~D'Amico, J.~Norena, and F.~Vernizzi, ``{The Effective Theory
  of Quintessence: the w<-1 Side Unveiled},'' {\em JCAP} {\bf 0902} (2009) 018,
  \href{http://xxx.lanl.gov/abs/0811.0827}{{\tt 0811.0827}}.

\bibitem{Arnowitt:1962hi}
R.~L. Arnowitt, S.~Deser, and C.~W. Misner, ``{The Dynamics of general
  relativity},'' {\em Gen.Rel.Grav.} {\bf 40} (2008) 1997--2027,
  \href{http://xxx.lanl.gov/abs/gr-qc/0405109}{{\tt gr-qc/0405109}}.

\bibitem{Bloomfield:2012ff}
J.~K. Bloomfield, {\'E}.~{\'E}. Flanagan, M.~Park, and S.~Watson, ``{Dark
  energy or modified gravity? An effective field theory approach},'' {\em JCAP}
  {\bf 1308} (2013) 010, \href{http://xxx.lanl.gov/abs/1211.7054}{{\tt
  1211.7054}}.

\bibitem{Bloomfield:2013efa}
J.~Bloomfield, ``{A Simplified Approach to General Scalar-Tensor Theories},''
  {\em JCAP} {\bf 1312} (2013) 044,
  \href{http://xxx.lanl.gov/abs/1304.6712}{{\tt 1304.6712}}.

\bibitem{Piazza:2013coa}
F.~Piazza and F.~Vernizzi, ``{Effective Field Theory of Cosmological
  Perturbations},'' {\em Class.Quant.Grav.} {\bf 30} (2013) 214007,
  \href{http://xxx.lanl.gov/abs/1307.4350}{{\tt 1307.4350}}.

\bibitem{Tsujikawa:2014mba}
S.~Tsujikawa, ``{The effective field theory of inflation/dark energy and the
  Horndeski theory},'' {\em Lect.Notes Phys.} {\bf 892} (2015) 97--136,
  \href{http://xxx.lanl.gov/abs/1404.2684}{{\tt 1404.2684}}.

\bibitem{Gleyzes:2014rba}
J.~Gleyzes, D.~Langlois, and F.~Vernizzi, ``{A unifying description of dark
  energy},'' {\em Int.J.Mod.Phys.} {\bf D23} (2014) 3010,
  \href{http://xxx.lanl.gov/abs/1411.3712}{{\tt 1411.3712}}.

\bibitem{Bloomfield:2013cyf}
J.~Bloomfield and J.~Pearson, ``{Simple implementation of general dark energy
  models},'' {\em JCAP} {\bf 1403} (2014) 017,
  \href{http://xxx.lanl.gov/abs/1310.6033}{{\tt 1310.6033}}.

\bibitem{Piazza:2013pua}
F.~Piazza, H.~Steigerwald, and C.~Marinoni, ``{Phenomenology of dark energy:
  exploring the space of theories with future redshift surveys},'' {\em JCAP}
  {\bf 1405} (2014) 043, \href{http://xxx.lanl.gov/abs/1312.6111}{{\tt
  1312.6111}}.

\bibitem{Bellini:2014fua}
E.~Bellini and I.~Sawicki, ``{Maximal freedom at minimum cost: linear
  large-scale structure in general modifications of gravity},'' {\em JCAP} {\bf
  1407} (2014) 050, \href{http://xxx.lanl.gov/abs/1404.3713}{{\tt 1404.3713}}.

\bibitem{Hu:2013twa}
B.~Hu, M.~Raveri, N.~Frusciante, and A.~Silvestri, ``{Effective Field Theory of
  Cosmic Acceleration: an implementation in CAMB},'' {\em Phys.Rev.} {\bf D89}
  (2014), no.~10 103530, \href{http://xxx.lanl.gov/abs/1312.5742}{{\tt
  1312.5742}}.

\bibitem{Ade:2015rim}
{\bf Planck} Collaboration, P.~Ade {\em et.~al.}, ``{Planck 2015 results. XIV.
  Dark energy and modified gravity},''
  \href{http://xxx.lanl.gov/abs/1502.01590}{{\tt 1502.01590}}.

\bibitem{Horndeski:1974wa}
G.~W. Horndeski, ``{Second-order scalar-tensor field equations in a
  four-dimensional space},'' {\em Int.J.Theor.Phys.} {\bf 10} (1974) 363--384.

\bibitem{Deffayet:2011gz}
C.~Deffayet, X.~Gao, D.~Steer, and G.~Zahariade, ``{From k-essence to
  generalised Galileons},'' {\em Phys.Rev.} {\bf D84} (2011) 064039,
  \href{http://xxx.lanl.gov/abs/1103.3260}{{\tt 1103.3260}}.

\bibitem{Kobayashi:2011nu}
T.~Kobayashi, M.~Yamaguchi, and J.~Yokoyama, ``{Generalized G-inflation:
  Inflation with the most general second-order field equations},'' {\em
  Prog.Theor.Phys.} {\bf 126} (2011) 511--529,
  \href{http://xxx.lanl.gov/abs/1105.5723}{{\tt 1105.5723}}.

\bibitem{Gleyzes:2014dya}
J.~Gleyzes, D.~Langlois, F.~Piazza, and F.~Vernizzi, ``{Healthy theories beyond
  Horndeski},'' \href{http://xxx.lanl.gov/abs/1404.6495}{{\tt 1404.6495}}.

\bibitem{Gleyzes:2014qga}
J.~Gleyzes, D.~Langlois, F.~Piazza, and F.~Vernizzi, ``{Exploring gravitational
  theories beyond Horndeski},'' {\em JCAP} {\bf 1502} (2015), no.~02 018,
  \href{http://xxx.lanl.gov/abs/1408.1952}{{\tt 1408.1952}}.

\bibitem{Zumalacarregui:2013pma}
M.~Zumalac{\'a}rregui and J.~Garc{\'\i}a-Bellido, ``{Transforming gravity: from
  derivative couplings to matter to second-order scalar-tensor theories beyond
  the Horndeski Lagrangian},'' {\em Phys.Rev.} {\bf D89} (2014), no.~6 064046,
  \href{http://xxx.lanl.gov/abs/1308.4685}{{\tt 1308.4685}}.

\bibitem{Will:2014xja}
C.~M. Will, ``{The Confrontation between General Relativity and Experiment},''
  {\em Living Rev.Rel.} {\bf 17} (2014) 4,
  \href{http://xxx.lanl.gov/abs/1403.7377}{{\tt 1403.7377}}.

\bibitem{Brans:1961sx}
C.~Brans and R.~Dicke, ``{Mach's principle and a relativistic theory of
  gravitation},'' {\em Phys.Rev.} {\bf 124} (1961) 925--935.

\bibitem{Hui:2009kc}
L.~Hui, A.~Nicolis, and C.~Stubbs, ``{Equivalence Principle Implications of
  Modified Gravity Models},'' {\em Phys.Rev.} {\bf D80} (2009) 104002,
  \href{http://xxx.lanl.gov/abs/0905.2966}{{\tt 0905.2966}}.

\bibitem{Creminelli:2013nua}
P.~Creminelli, J.~Gleyzes, L.~Hui, M.~Simonovi{\'c}, and F.~Vernizzi,
  ``{Single-Field Consistency Relations of Large Scale Structure. Part III:
  Test of the Equivalence Principle},'' {\em JCAP} {\bf 1406} (2014) 009,
  \href{http://xxx.lanl.gov/abs/1312.6074}{{\tt 1312.6074}}.

\bibitem{Damour:1990tw}
T.~Damour, G.~W. Gibbons, and C.~Gundlach, ``{Dark Matter, Time Varying $G$,
  and a Dilaton Field},'' {\em Phys.Rev.Lett.} {\bf 64} (1990) 123--126.

\bibitem{Amendola:1999er}
L.~Amendola, ``{Coupled quintessence},'' {\em Phys.Rev.} {\bf D62} (2000)
  043511, \href{http://xxx.lanl.gov/abs/astro-ph/9908023}{{\tt
  astro-ph/9908023}}.

\bibitem{Skordis:2015yra}
C.~Skordis, A.~Pourtsidou, and E.~Copeland, ``{The Parameterized
  Post-Friedmannian Framework for Interacting Dark Energy Theories},''
  \href{http://xxx.lanl.gov/abs/1502.07297}{{\tt 1502.07297}}.

\bibitem{Baker:2012zs}
T.~Baker, P.~G. Ferreira, and C.~Skordis, ``{The Parameterized Post-Friedmann
  framework for theories of modified gravity: concepts, formalism and
  examples},'' {\em Phys.Rev.} {\bf D87} (2013), no.~2 024015,
  \href{http://xxx.lanl.gov/abs/1209.2117}{{\tt 1209.2117}}.

\bibitem{Ferreira:2014mja}
P.~G. Ferreira, T.~Baker, and C.~Skordis, ``{Testing general relativity with
  cosmology: a synopsis of the parametrized post-Friedmann approach},'' {\em
  Gen.Rel.Grav.} {\bf 46} (2014) 1788.

\bibitem{Bettoni:2013diz}
D.~Bettoni and S.~Liberati, ``{Disformal invariance of second order
  scalar-tensor theories: Framing the Horndeski action},'' {\em Phys.Rev.} {\bf
  D88} (2013), no.~8 084020, \href{http://xxx.lanl.gov/abs/1306.6724}{{\tt
  1306.6724}}.

\bibitem{Bekenstein:1992pj}
J.~D. Bekenstein, ``{The Relation between physical and gravitational
  geometry},'' {\em Phys.Rev.} {\bf D48} (1993) 3641--3647,
  \href{http://xxx.lanl.gov/abs/gr-qc/9211017}{{\tt gr-qc/9211017}}.

\bibitem{Blas:2012vn}
D.~Blas, M.~M. Ivanov, and S.~Sibiryakov, ``{Testing Lorentz invariance of dark
  matter},'' {\em JCAP} {\bf 1210} (2012) 057,
  \href{http://xxx.lanl.gov/abs/1209.0464}{{\tt 1209.0464}}.

\bibitem{Bettoni:2015wla}
D.~Bettoni and S.~Liberati, ``{Dynamics of non-minimally coupled perfect
  fluids},'' \href{http://xxx.lanl.gov/abs/1502.06613}{{\tt 1502.06613}}.

\bibitem{Koivisto:2008ak}
T.~S. Koivisto, ``{Disformal quintessence},''
  \href{http://xxx.lanl.gov/abs/0811.1957}{{\tt 0811.1957}}.

\bibitem{Zumalacarregui:2010wj}
M.~Zumalac\'arregui, T.~Koivisto, D.~Mota, and P.~Ruiz-Lapuente, ``{Disformal
  Scalar Fields and the Dark Sector of the Universe},'' {\em JCAP} {\bf 1005}
  (2010) 038, \href{http://xxx.lanl.gov/abs/1004.2684}{{\tt 1004.2684}}.

\bibitem{Koivisto:2012za}
T.~S. Koivisto, D.~F. Mota, and M.~Zumalac\'arregui, ``{Screening Modifications
  of Gravity through Disformally Coupled Fields},'' {\em Phys.Rev.Lett.} {\bf
  109} (2012) 241102, \href{http://xxx.lanl.gov/abs/1205.3167}{{\tt
  1205.3167}}.

\bibitem{vandeBruck:2012vq}
C.~van~de Bruck and G.~Sculthorpe, ``{Modified Gravity and the Radiation
  Dominated Epoch},'' {\em Phys.Rev.} {\bf D87} (2013), no.~4 044004,
  \href{http://xxx.lanl.gov/abs/1210.2168}{{\tt 1210.2168}}.

\bibitem{Zumalacarregui:2012us}
M.~Zumalac\'arregui, T.~S. Koivisto, and D.~F. Mota, ``{DBI Galileons in the
  Einstein Frame: Local Gravity and Cosmology},'' {\em Phys.Rev.} {\bf D87}
  (2013) 083010, \href{http://xxx.lanl.gov/abs/1210.8016}{{\tt 1210.8016}}.

\bibitem{Brax:2013nsa}
P.~Brax, C.~Burrage, A.-C. Davis, and G.~Gubitosi, ``{Cosmological Tests of the
  Disformal Coupling to Radiation},'' {\em JCAP} {\bf 1311} (2013) 001,
  \href{http://xxx.lanl.gov/abs/1306.4168}{{\tt 1306.4168}}.

\bibitem{Brax:2014vva}
P.~Brax and C.~Burrage, ``{Constraining Disformally Coupled Scalar Fields},''
  {\em Phys.Rev.} {\bf D90} (2014), no.~10 104009,
  \href{http://xxx.lanl.gov/abs/1407.1861}{{\tt 1407.1861}}.

\bibitem{Sakstein:2014isa}
J.~Sakstein, ``{Disformal Theories of Gravity: From the Solar System to
  Cosmology},'' {\em JCAP} {\bf 1412} (2014), no.~12 012,
  \href{http://xxx.lanl.gov/abs/1409.1734}{{\tt 1409.1734}}.

\bibitem{vandeBruck:2015ida}
C.~van~de Bruck and J.~Morrice, ``{Disformal couplings and the dark sector of
  the universe},'' \href{http://xxx.lanl.gov/abs/1501.03073}{{\tt 1501.03073}}.

\bibitem{Koivisto:2015mwa}
T.~Koivisto and H.~J. Nyrhinen, ``{Stability of disformally coupled accretion
  disks},'' \href{http://xxx.lanl.gov/abs/1503.02063}{{\tt 1503.02063}}.

\bibitem{Billyard:2000bh}
A.~P. Billyard and A.~A. Coley, ``{Interactions in scalar field cosmology},''
  {\em Phys.Rev.} {\bf D61} (2000) 083503,
  \href{http://xxx.lanl.gov/abs/astro-ph/9908224}{{\tt astro-ph/9908224}}.

\bibitem{Farrar:2003uw}
G.~R. Farrar and P.~J.~E. Peebles, ``{Interacting dark matter and dark
  energy},'' {\em Astrophys.J.} {\bf 604} (2004) 1--11,
  \href{http://xxx.lanl.gov/abs/astro-ph/0307316}{{\tt astro-ph/0307316}}.

\bibitem{Amendola:2003wa}
L.~Amendola, ``{Linear and non-linear perturbations in dark energy models},''
  {\em Phys.Rev.} {\bf D69} (2004) 103524,
  \href{http://xxx.lanl.gov/abs/astro-ph/0311175}{{\tt astro-ph/0311175}}.

\bibitem{Bertolami:2007zm}
O.~Bertolami, F.~Gil~Pedro, and M.~Le~Delliou, ``{Dark Energy-Dark Matter
  Interaction and the Violation of the Equivalence Principle from the Abell
  Cluster A586},'' {\em Phys.Lett.} {\bf B654} (2007) 165--169,
  \href{http://xxx.lanl.gov/abs/astro-ph/0703462}{{\tt astro-ph/0703462}}.

\bibitem{Baldi:2008ay}
M.~Baldi, V.~Pettorino, G.~Robbers, and V.~Springel, ``{Hydrodynamical N-body
  simulations of coupled dark energy cosmologies},'' {\em
  Mon.Not.Roy.Astron.Soc.} {\bf 403} (2010) 1684--1702,
  \href{http://xxx.lanl.gov/abs/0812.3901}{{\tt 0812.3901}}.

\bibitem{Koyama:2009gd}
K.~Koyama, R.~Maartens, and Y.-S. Song, ``{Velocities as a probe of dark sector
  interactions},'' {\em JCAP} {\bf 0910} (2009) 017,
  \href{http://xxx.lanl.gov/abs/0907.2126}{{\tt 0907.2126}}.

\bibitem{Valiviita:2009nu}
J.~Valiviita, R.~Maartens, and E.~Majerotto, ``{Observational constraints on an
  interacting dark energy model},'' {\em Mon.Not.Roy.Astron.Soc.} {\bf 402}
  (2010) 2355--2368, \href{http://xxx.lanl.gov/abs/0907.4987}{{\tt 0907.4987}}.

\bibitem{Baldi:2010vv}
M.~Baldi, ``{Time dependent couplings in the dark sector: from background
  evolution to nonlinear structure formation},'' {\em Mon.Not.Roy.Astron.Soc.}
  {\bf 411} (2011) 1077, \href{http://xxx.lanl.gov/abs/1005.2188}{{\tt
  1005.2188}}.

\bibitem{Pettorino:2012ts}
V.~Pettorino, L.~Amendola, C.~Baccigalupi, and C.~Quercellini, ``{Constraints
  on coupled dark energy using CMB data from WMAP and SPT},'' {\em Phys.Rev.}
  {\bf D86} (2012) 103507, \href{http://xxx.lanl.gov/abs/1207.3293}{{\tt
  1207.3293}}.

\bibitem{yang:2014vza}
W.~Yang and L.~Xu, ``{Testing coupled dark energy with large scale structure
  observation},'' {\em JCAP} {\bf 1408} (2014) 034,
  \href{http://xxx.lanl.gov/abs/1401.5177}{{\tt 1401.5177}}.

\bibitem{Salvatelli:2014zta}
V.~Salvatelli, N.~Said, M.~Bruni, A.~Melchiorri, and D.~Wands, ``{Indications
  of a late-time interaction in the dark sector},'' {\em Phys.Rev.Lett.} {\bf
  113} (2014), no.~18 181301, \href{http://xxx.lanl.gov/abs/1406.7297}{{\tt
  1406.7297}}.

\bibitem{Bruneton:2007si}
J.-P. Bruneton and G.~Esposito-Farese, ``{Field-theoretical formulations of
  MOND-like gravity},'' {\em Phys.Rev.} {\bf D76} (2007) 124012,
  \href{http://xxx.lanl.gov/abs/0705.4043}{{\tt 0705.4043}}.

\bibitem{Hawking:1973uf}
S.~Hawking and G.~Ellis, {\em {The Large scale structure of space-time}}.
\newblock Cambridge University Press, 1975.

\bibitem{Creminelli:2014wna}
P.~Creminelli, J.~Gleyzes, J.~Nore{\~n}a, and F.~Vernizzi, ``{Resilience of the
  standard predictions for primordial tensor modes},'' {\em Phys.Rev.Lett.}
  {\bf 113} (2014), no.~23 231301,
  \href{http://xxx.lanl.gov/abs/1407.8439}{{\tt 1407.8439}}.

\bibitem{Malik:2004tf}
K.~A. Malik and D.~Wands, ``{Adiabatic and entropy perturbations with
  interacting fluids and fields},'' {\em JCAP} {\bf 0502} (2005) 007,
  \href{http://xxx.lanl.gov/abs/astro-ph/0411703}{{\tt astro-ph/0411703}}.

\bibitem{Minamitsuji:2014waa}
M.~Minamitsuji, ``{Disformal transformation of cosmological perturbations},''
  {\em Phys.Lett.} {\bf B737} (2014) 139--150,
  \href{http://xxx.lanl.gov/abs/1409.1566}{{\tt 1409.1566}}.

\bibitem{Weinberg:2003sw}
S.~Weinberg, ``{Adiabatic modes in cosmology},'' {\em Phys.Rev.} {\bf D67}
  (2003) 123504, \href{http://xxx.lanl.gov/abs/astro-ph/0302326}{{\tt
  astro-ph/0302326}}.

\bibitem{Bardeen:1983qw}
J.~M. Bardeen, P.~J. Steinhardt, and M.~S. Turner, ``{Spontaneous Creation of
  Almost Scale - Free Density Perturbations in an Inflationary Universe},''
  {\em Phys.Rev.} {\bf D28} (1983) 679.

\bibitem{Dupays:2006dp}
A.~Dupays, E.~Masso, J.~Redondo, and C.~Rizzo, ``{Light scalars coupled to
  photons and non-newtonian forces},'' {\em Phys.Rev.Lett.} {\bf 98} (2007)
  131802, \href{http://xxx.lanl.gov/abs/hep-ph/0610286}{{\tt hep-ph/0610286}}.

\bibitem{Fardon:2003eh}
R.~Fardon, A.~E. Nelson, and N.~Weiner, ``{Dark energy from mass varying
  neutrinos},'' {\em JCAP} {\bf 0410} (2004) 005,
  \href{http://xxx.lanl.gov/abs/astro-ph/0309800}{{\tt astro-ph/0309800}}.

\bibitem{Brookfield:2005td}
A.~Brookfield, C.~van~de Bruck, D.~Mota, and D.~Tocchini-Valentini,
  ``{Cosmology with massive neutrinos coupled to dark energy},'' {\em
  Phys.Rev.Lett.} {\bf 96} (2006) 061301,
  \href{http://xxx.lanl.gov/abs/astro-ph/0503349}{{\tt astro-ph/0503349}}.

\bibitem{Afshordi:2005ym}
N.~Afshordi, M.~Zaldarriaga, and K.~Kohri, ``{On the stability of dark energy
  with mass-varying neutrinos},'' {\em Phys.Rev.} {\bf D72} (2005) 065024,
  \href{http://xxx.lanl.gov/abs/astro-ph/0506663}{{\tt astro-ph/0506663}}.

\bibitem{Sawicki:2015zya}
I.~Sawicki and E.~Bellini, ``{Limits of Quasi-Static Approximation in
  Modified-Gravity Cosmologies},''
  \href{http://xxx.lanl.gov/abs/1503.06831}{{\tt 1503.06831}}.

\bibitem{Bean:2003fb}
R.~Bean and O.~Dore, ``{Probing dark energy perturbations: The Dark energy
  equation of state and speed of sound as measured by WMAP},'' {\em Phys.Rev.}
  {\bf D69} (2004) 083503, \href{http://xxx.lanl.gov/abs/astro-ph/0307100}{{\tt
  astro-ph/0307100}}.

\bibitem{ArmendarizPicon:2000dh}
C.~Armendariz-Picon, V.~F. Mukhanov, and P.~J. Steinhardt, ``{A Dynamical
  solution to the problem of a small cosmological constant and late time cosmic
  acceleration},'' {\em Phys.Rev.Lett.} {\bf 85} (2000) 4438--4441,
  \href{http://xxx.lanl.gov/abs/astro-ph/0004134}{{\tt astro-ph/0004134}}.

\bibitem{ArmendarizPicon:2000ah}
C.~Armendariz-Picon, V.~F. Mukhanov, and P.~J. Steinhardt, ``{Essentials of k
  essence},'' {\em Phys.Rev.} {\bf D63} (2001) 103510,
  \href{http://xxx.lanl.gov/abs/astro-ph/0006373}{{\tt astro-ph/0006373}}.

\bibitem{Boubekeur:2008kn}
L.~Boubekeur, P.~Creminelli, J.~Norena, and F.~Vernizzi, ``{Action approach to
  cosmological perturbations: the 2nd order metric in matter dominance},'' {\em
  JCAP} {\bf 0808} (2008) 028, \href{http://xxx.lanl.gov/abs/0806.1016}{{\tt
  0806.1016}}.

\bibitem{Garriga:1999vw}
J.~Garriga and V.~F. Mukhanov, ``{Perturbations in k-inflation},'' {\em
  Phys.Lett.} {\bf B458} (1999) 219--225,
  \href{http://xxx.lanl.gov/abs/hep-th/9904176}{{\tt hep-th/9904176}}.

\bibitem{Gergely:2014rna}
L.~{\'A}. Gergely and S.~Tsujikawa, ``{Effective field theory of modified
  gravity with two scalar fields: dark energy and dark matter},'' {\em
  Phys.Rev.} {\bf D89} (2014), no.~6 064059,
  \href{http://xxx.lanl.gov/abs/1402.0553}{{\tt 1402.0553}}.

\bibitem{Maldacena:2002vr}
J.~M. Maldacena, ``{Non-Gaussian features of primordial fluctuations in single
  field inflationary models},'' {\em JHEP} {\bf 0305} (2003) 013,
  \href{http://xxx.lanl.gov/abs/astro-ph/0210603}{{\tt astro-ph/0210603}}.

\bibitem{Ma:1995ey}
C.-P. Ma and E.~Bertschinger, ``{Cosmological perturbation theory in the
  synchronous and conformal Newtonian gauges},'' {\em Astrophys.J.} {\bf 455}
  (1995) 7--25, \href{http://xxx.lanl.gov/abs/astro-ph/9506072}{{\tt
  astro-ph/9506072}}.

\end{thebibliography}\endgroup

\end{document}